\documentclass[12pt]{article}
\usepackage{a4wide,epsfig,psfrag,amsmath,amssymb,cite,scalefnt}
\usepackage{color}
\usepackage{comment}
\usepackage{epstopdf}

\graphicspath{ {figs/} }

\allowdisplaybreaks

\begin{document}

\title{\vskip-3cm{\baselineskip14pt
    \begin{flushleft}
    \normalsize TTP17-026
    \end{flushleft}} \vskip1.5cm 
The $n_f^2$ contributions to fermionic\\ four-loop form factors
}

\author{
Roman N. Lee$^{a}$,
Alexander V. Smirnov$^{b}$,
\\
Vladimir A. Smirnov$^{c}$,
and Matthias Steinhauser$^{d}$
\\[1em]
{\small\it (a) Budker Institute of Nuclear Physics,}\\
{\small\it 630090 Novosibirsk, Russia}
\\
{\small\it (b) Research Computing Center, Moscow State University}\\
{\small\it 119991, Moscow, Russia}
\\  
{\small\it (c) Skobeltsyn Institute of Nuclear Physics of Moscow State University}\\
{\small\it 119991, Moscow, Russia}
\\
{\small\it (d) Institut f{\"u}r Theoretische Teilchenphysik,
 Karlsruhe Institute of Technology (KIT)}\\
 {\small\it 76128 Karlsruhe, Germany}  
}
  
\date{}

\maketitle

\thispagestyle{empty}

\begin{abstract}
  We compute the four-loop contributions to the photon quark and Higgs
  quark form factors involving two closed fermion loops.  We present
  analytical results for all non-planar master integrals of the two non-planar
  integral families which enter our calculation.

  \medskip


\end{abstract}

\thispagestyle{empty}


\newpage


\section{Introduction}

There are various aspects of form factors which promote them to important
objects in any quantum field theory. In the framework of QCD form factors
constitute building blocks for various production and decay processes, most
prominently for Higgs boson production and the Drell-Yan process.
Furthermore, form factors are the simplest Green's functions with a non-trivial
infrared structure which makes them ideal objects to investigate general
infrared properties of gauge theories~\cite{Mueller:1979ih,Collins:1980ih,Sen:1981sd,Magnea:1990zb,Korchemsky:1991zp,Korchemskaya:1992je,Sterman:2002qn}.

The main objects of this work are massless fermionic form factors where the
fermions couple via a vector and scalar coupling to an external current. In
the framework of the Standard Model they can be interpreted as photon quark
and Higgs quark form factors. In the following we provide brief definitions of
these objects.

The quark-antiquark-photon form factor is conveniently obtained from the
photon quark vertex function $\Gamma^{\mu}_q$ by applying an appropriate
projector. In $d=4-2\epsilon$ space-time dimensions we have
\begin{eqnarray}
  F_q(q^2) &=& -\frac{1}{4(1-\epsilon)q^2}
  \mbox{Tr}\left( q_2\!\!\!\!\!/\,\,\, \Gamma^\mu_q \, q_1\!\!\!\!\!/\,\,\,
  \gamma_\mu\right)
  \,,
\end{eqnarray}
with $q=q_1+q_2$. $q_1$ and $q_2$ are the incoming quark and antiquark
momenta and $q$ is the momentum of the photon.

\begin{figure}[b] 
  \begin{center}
    \includegraphics[width=0.9\textwidth]{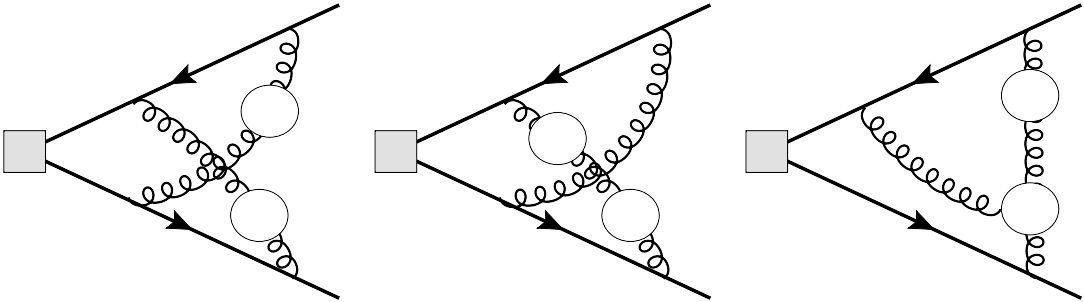}
    \caption{\label{fig::diags}Sample Feynman diagrams contributing to $F_q$
      and $F_b$ at the four-loop order and containing two closed fermion
      loops.
      The gray box indicates either a
      scalar or a vector coupling of the fermions to the external current.
      Solid and curly lines represent quarks and gluons, respectively. All
      particles are massless.}
  \end{center}
\end{figure}

In analogy, the Higgs quark form factor is constructed from the Higgs quark
vertex function $\Gamma_q$. For definiteness we consider in the following the
coupling to bottom quarks and write
\begin{eqnarray}
  F_b(q^2) &=& -\frac{1}{2q^2}
  \mbox{Tr}\left( q_2\!\!\!\!\!/\,\,\, \Gamma_b \, q_1\!\!\!\!\!/\,\,\,
  \right)
  \,.
\end{eqnarray}
Sample Feynman diagrams contributing to $F_q$ and $F_b$ are shown in
Fig.~\ref{fig::diags}. 

Two- and three-loop corrections to $F_q$ have been computed in
Refs.~\cite{Kramer:1986sg,Matsuura:1987wt,Matsuura:1988sm,Gehrmann:2005pd,Baikov:2009bg,Gehrmann:2010ue,Lee:2010ik,Gehrmann:2010tu}
and $F_b$ has been considered in
Refs.~\cite{Anastasiou:2011qx,Gehrmann:2014vha}.  Four-loop results for $F_q$
in the planar limit have been obtained in
Refs.~\cite{Henn:2016men,Lee:2016ixa} and fermionic corrections with three
closed quark loops have been computed in Ref.~\cite{vonManteuffel:2016xki}.

In this paper we consider fermionic contributions to $F_q$ and $F_b$.  More
precisely, we compute four-loop corrections with two closed fermion loops
(which in our notation are proportional to $n_f^2$).  This is the simplest
well-defined gauge invariant subset which involves non-planar integral
families. It is the main result of this work to study in detail these families
and provide analytic results for all non-planar master integrals which are
part of these families.  Note, that all planar families including all master
integrals have been considered in a systematic way in
Refs.~\cite{Henn:2016men,Lee:2016ixa,HSS}.  Non-planar integral families are
already present in the $n_f^3$ contribution to the Higgs gluon form factor
which has been considered in Ref.~\cite{vonManteuffel:2016xki}.  Very
recently the $1/\epsilon^2$ pole of the four-loop form factor within 
${\cal N}=4$ super Yang-Mills theory has been computed using numerical
methods~\cite{Boels:2017skl}.

If $\epsilon^0$ terms at four loops shall be computed the lower-order
corrections need to be expanded to higher orders in $\epsilon$. In particular,
the one-, two- and three-loop results are needed to order $\epsilon^6$,
$\epsilon^4$ and $\epsilon^2$, respectively. The three-loop $\epsilon^2$ terms
for $F_q$ have been computed in Ref.~\cite{Gehrmann:2010ue} and 
cross-checked in Ref.~\cite{Lee:2016ixa}. As a preparatory calculation for the
results obtained in this work we could confirm the three-loop corrections to
the gluonic from factor up to order $\epsilon^2$ as given in
Ref.~\cite{Gehrmann:2010ue}. Furthermore, we provide the first independent
check for the three-loop $\epsilon^0$ term of $F_b$~\cite{Gehrmann:2014vha}
and extend it to order $\epsilon^2$ which is not yet available in the
literature.

It is convenient to parametrize the form factor in terms of the bare strong
coupling constant and write
\begin{eqnarray}
  F_q &=& 1 +
  \sum_{n\ge1} 
  \left(\frac{\alpha_s^0}{4\pi}\right)^n
  \left(\frac{\mu^2}{-q^2-i0} \right)^{n\epsilon}
  F_q^{(n)}
  \,,
  \nonumber\\
  F_b &=& y^0\left[1 +
  \sum_{n\ge1} 
  \left(\frac{\alpha_s^0}{4\pi}\right)^n
  \left(\frac{\mu^2}{-q^2-i0} \right)^{n\epsilon}
  F_b^{(n)}
  \right]
  \,,
  \label{eq::FFbare}
\end{eqnarray}
where $y^0=m_b^0/v$ is the bare Yukawa coupling and $m_b^0$
and $v$ are the bare bottom quark mass and Higgs vacuum expectation value,
respectively. The renormalized counterparts of the form factors are easily obtained by
multiplicative renormalization of the parameters $\alpha_s^0$ and $y^0$ using
three- and four-loop renormalization constants, respectively, which can be
obtained from
Refs.~\cite{Tarasov:1980au,Larin:1993tp,Chetyrkin:1997dh,Vermaseren:1997fq}
(see also Appendix~F and the ancillary file of Ref.~\cite{Marquard:2016dcn}
for an explicit result of $Z_y$ up to four loops in terms of SU$(N_c)$ colour factors).

The pole part of the logarithm of the renormalized form factor
has a universal structure which is given by
(see, e.g., Ref.~\cite{Korchemsky:1987wg,Becher:2009qa,Gehrmann:2010ue}) 
\newcommand{\ci}{C_F}
\begin{eqnarray}
\lefteqn{  \log( F_x ) =}
\nonumber\\&&\mbox{}
\frac{\alpha_s}{4\pi}  \Bigg\{
\frac{1}{\epsilon^2}  \Bigg[
-\frac{1}{2} \ci \gamma _{\text{cusp}}^0
\Bigg]
+\frac{1}{\epsilon}  \Bigg[
\gamma _x^0
\Bigg]
\Bigg\}
\nonumber\\&&\mbox{}
+\left(\frac{\alpha_s}{4\pi}\right)^2  \Bigg\{
\frac{1}{\epsilon^3}  \Bigg[
\frac{3}{8} \beta _0 \ci \gamma _{\text{cusp}}^0
\Bigg]
+\frac{1}{\epsilon^2}  \Bigg[
-\frac{1}{2} \beta _0 \gamma _x^0-\frac{1}{8} \ci \gamma _{\text{cusp}}^1
\Bigg]
+\frac{1}{\epsilon}  \Bigg[
\frac{\gamma _x^1}{2}
\Bigg]
\Bigg\}
\nonumber\\&&\mbox{}
+\left(\frac{\alpha_s}{4\pi}\right)^3  \Bigg\{
\frac{1}{\epsilon^4}  \Bigg[
-\frac{11}{36} \beta _0^2 \ci \gamma _{\text{cusp}}^0
\Bigg]
+\frac{1}{\epsilon^3}  \Bigg[
\ci \left(\frac{2}{9} \beta _1 \gamma _{\text{cusp}}^0+\frac{5}{36} \beta _0 \gamma _{\text{cusp}}^1\right)+\frac{1}{3} \beta _0^2 \gamma _x^0
\Bigg]
\nonumber\\&&\mbox{}
+\frac{1}{\epsilon^2}  \Bigg[
-\frac{1}{3} \beta _1 \gamma _x^0-\frac{1}{3} \beta _0 \gamma _x^1-\frac{1}{18} \ci \gamma _{\text{cusp}}^2
\Bigg]
+\frac{1}{\epsilon}  \Bigg[
\frac{\gamma _x^2}{3}
\Bigg]
\Bigg\}
\nonumber\\&&\mbox{}
+\left(\frac{\alpha_s}{4\pi}\right)^4  \Bigg\{
\frac{1}{\epsilon^5}  \Bigg[
\frac{25}{96} \beta _0^3 \ci \gamma _{\text{cusp}}^0
\Bigg]
+\frac{1}{\epsilon^4}  \Bigg[
\ci \left(-\frac{13}{96} \beta _0^2 \gamma _{\text{cusp}}^1-\frac{5}{12} \beta _1 \beta _0 \gamma _{\text{cusp}}^0\right)-\frac{1}{4} \beta _0^3 \gamma _x^0
\Bigg]
\nonumber\\&&\mbox{}
+\frac{1}{\epsilon^3}  \Bigg[
\ci \left(\frac{5}{32} \beta _2 \gamma _{\text{cusp}}^0+\frac{3}{32} \beta _1 \gamma _{\text{cusp}}^1+\frac{7}{96} \beta _0 \gamma _{\text{cusp}}^2\right)+\frac{1}{4} \beta _0^2 \gamma _x^1+\frac{1}{2} \beta _1 \beta _0 \gamma _x^0
\Bigg]
\nonumber\\&&\mbox{}
+\frac{1}{\epsilon^2}  \Bigg[
-\frac{1}{4} \beta _2 \gamma _x^0-\frac{1}{4} \beta _1 \gamma _x^1-\frac{1}{4} \beta _0 \gamma _x^2-\frac{1}{32} \ci \gamma _{\text{cusp}}^3
\Bigg]
+\frac{1}{\epsilon}  \Bigg[
\frac{\gamma _x^3}{4}
\Bigg]
\Bigg\}
+ \ldots
\,,
  \label{eq::logFx_gen}
\end{eqnarray}
where the ellipse denote higher order terms in $\alpha_s$ and $\epsilon$.  We
have $x\in\{q,b\}$.
$C_F$ and $C_A$ are the eigenvalues of the quadratic Casimir operators of the
fundamental and adjoint representation for the SU$(N_c)$ colour group,
respectively.  In Eq.~(\ref{eq::logFx_gen}) $\mu^2=-q^2$ has been
chosen. The cusp and collinear
anomalous dimensions are defined through
\begin{eqnarray}
  \gamma_x &=& \sum_{n\ge0} \left(\frac{\alpha_s{(\mu^2)}}{4\pi}\right)^n
  \gamma_x^n
  \,,
  \label{eq::gamma_x}
\end{eqnarray}
with $x\in\{\text{cusp},q,b\}$.
Note that $\gamma_q = \gamma_b$ which, together with the universality of
$\gamma_{\rm cusp}$, is an important check of our calculation.
The coefficients of the $\beta$ function in Eq.~(\ref{eq::logFx_gen})
are given by
\begin{eqnarray}
  \beta_0 &=& \frac{11 C_A}{3}-\frac{2 n_f}{3}
  \,,\nonumber\\
  \beta_1 &=& -\frac{10 C_A n_f}{3}+\frac{34 C_A^2}{3}-2 C_F n_f
  \,,\nonumber\\
  \beta_2 &=& 
   -\frac{205}{18} C_A C_F n_f-\frac{1415}{54} C_A^2 n_f+\frac{79}{54}   C_A
    n_f^2+\frac{2857 C_A^3}{54}+\frac{11}{9} C_F n_f^2+C_F^2 n_f
  \,.
\end{eqnarray}
where we have used $T=1/2$ with $T$ being the index of the fundamental
representation. $n_f$ counts the number of massless active quarks.

Note that the three leading pole terms of the four-loop contribution to
$\log(F_x)$ are determined from lower-order coefficients and serve as an
important consistency check. The $1/\epsilon^2$ and $1/\epsilon^1$ terms have
new ingredients, $\gamma_{\rm cusp}^3$ and $\gamma_x^3$, which we can extract
by comparison of Eq.~(\ref{eq::logFx_gen}) with our explicit calculation of
the form factor.

The remainder of the paper is organized as follows: In the next section we
provide details to the techniques which have been used to compute
non-planar four-loop vertex integrals.  Results for the $n_f^2$ contributions
are presented in Section~\ref{sec::res}. $F_q$ and $F_b$ are
discussed in Sections~\ref{sub::qff} and~\ref{sub::hbb}, respectively. In
particular, we provide results for the $n_f^2$ terms of the four-loop cusp and
collinear anomalous dimensions and the finite parts of the form factors.
Section~\ref{sec::concl} contains our conclusions.
In the Appendix we provide analytic results for all non-planar master
integrals which are present in the non-planar integral families needed for our
calculation. The analytic results presented in this paper are also
provided in computer-readable form in an ancillary file~\cite{progdata}.


\section{Technical details}

\begin{figure}[t] 
  \begin{center}
    \includegraphics[width=0.9\textwidth]{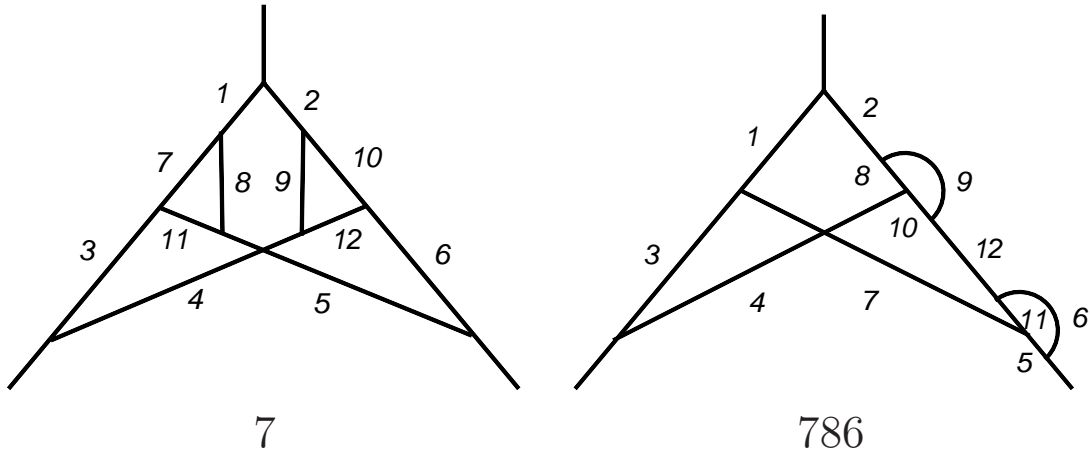}
    \caption{\label{fig::diags_fam}Graphs associated with the non-planar
      families 7 and 786 needed for the $n_f^2$ contribution to $F_q$ and $F_b$.  The
      numbers $n$ next to the lines correspond to the indices of the
      propagators, i.e. to the $n^{\rm th}$ integer argument of the functions
      representing the integrals. In addition to the 12 propagators we have
      for each family six linear independent numerator factors. However, the
      corresponding indices are always zero for our master integrals.  }
  \end{center}
\end{figure}

The amplitudes, which contribute to the form factors, are prepared with the help
of a well-tested setup. In a first step they are generated with {\tt
  qgraf}~\cite{Nogueira:1991ex}.  For the fermionic form factors we have in
total 1, 15 and 337 diagrams at one, two and three loops. At four-loop order
we have 77 diagrams proportional to $n_f^2$ and one diagram proportional to
$n_f^3$ (cf. Fig.~\ref{fig::diags} for sample diagrams).    Next,
we transform the output to {\tt FORM}~\cite{Kuipers:2012rf} notation using
{\tt q2e} and {\tt exp}~\cite{Harlander:1997zb,Seidensticker:1999bb}.  The
program {\tt exp} furthermore maps each Feynman diagram to families for
massless four-loop vertices with two different non-vanishing external momenta.
Then we perform the Dirac algebra and obtain a set of input integrals for each
family.  It turns out that fourteen planar and two
non-planar families are involved in the $n_f^2$ contribution of the
fermionic form factors.  The graphs associated with the non-planar
families are shown in Fig.~\ref{fig::diags_fam}.\footnote{For convenience we
  use the internal numeration of the families also in the paper.}

For the reduction to master integrals we use {\tt
  FIRE}~\cite{Smirnov:2008iw,Smirnov:2013dia,Smirnov:2014hma} which we apply
in combination with {\tt LiteRed}~\cite{Lee:2012cn,Lee:2013mka}.  Using {\tt
  FIRE} we reveal 26 and 40 one-scale master integrals for the two non-planar
families 7 and 786, respectively.  For the analytical computation of these
master integrals we follow the same strategy as in our previous
work~\cite{Lee:2016ixa} which we briefly summarize for convenience:
\begin{enumerate}
\item We introduce a second mass scale by removing one of the quark momenta,
  $q_2$, from the light cone, i.e., we have $q_2^2\not=0$. Furthermore we define
  $q_2^2 = x q^2$. With the help of {\tt FIRE} we obtain 91 (101) two-scale
  master integrals for family 7 (786).  The differential
  equations for these master integrals with respect to $x$ are obtained with
  the help of {\tt LiteRed}~\cite{Lee:2012cn,Lee:2013mka}.
\item To solve our differential equations we turn from the primary basis to
  a canonical basis~\cite{Henn:2013pwa}, where the corresponding master
  integrals satisfy a system of differential equations with the right-hand side
  proportional to $\epsilon$ and with only so-called Fuchsian singularities
  with respect to $x$.  To construct our canonical basis we apply the private
  implementation of one of the authors (R.N.L.) of the algorithm\footnote{Two
    public implementations, {\tt
      Fuchsia}~\cite{Gituliar:2016vfa,Gituliar:2017vzm} and {\tt
      epsilon}~\cite{Prausa:2017ltv}, of the algorithm of
    Ref.~\cite{Lee:2014ioa} are available. See also~\cite{Meyer:2016slj,Meyer:2017joq} where
    an algorithm for the case of two and more variables is described.}
  discussed in Ref.~\cite{Lee:2014ioa}.
\item Since our equations are in a canonical (or $\epsilon$) form, we write
  down a solution in a straightforward way order-by-order in $\epsilon$ in
  terms of harmonic polylogarithms (HPL)~\cite{Remiddi:1999ew} with letters 0
  and 1.
\item We determine the boundary conditions for the canonical master
  integrals, which are given as linear combinations of the primary master
  integrals, for $x=1$.  The primary master integrals are regular at
  this point  and become of propagator type. 
  Thus they are expressed as linear combinations of 28
  master integrals. Their analytic $\epsilon$-expansions are
  well-known~\cite{Baikov:2010hf,Lee:2011jt} up to
  weight 12. They have been cross-checked numerically in
  Ref.~\cite{Smirnov:2010hd}.
\item We solve our differential equations asymptotically near the point $x=0$
  and fix these solutions by matching them to our solution at general $x$.
  Here we use the package {\tt HPL}~\cite{Maitre:2005uu} to extract the
  leading order behaviour of the elements of the canonical basis in the limit
  $x\to 0$.  The asymptotic solutions are linear combinations of powers $x^{k
    \epsilon}$ with $k=0,1,\ldots,8$.  We pick up asymptotic terms with $k=0$
  and obtain the so-called naive values of the canonical master integrals at
  $x=0$.
\item From the analytic results for the naive part we obtain analytical
  results for the sought-after one-scale master integrals after changing back
  to the primary basis.
\end{enumerate}

To make the transition from the point $x=1$ to the point $x=0$ (cf. items
4. and 5. in the above list) we could apply the prescriptions explained in
our previous paper~\cite{Lee:2016ixa}. However, we prefer to use the
following slightly modified approach which we find more effective.

Let us assume that we have a differential equation in $\epsilon$-form
\begin{equation}
  \partial_x \mathbf{J}(x) = \epsilon M(x) \mathbf{J}(x)\,,\quad 
  M(x) = \sum_a \frac{M_a}{x-a}\,.
\end{equation}
In our case, the sum over $a$ includes two terms, with $a=0$ and  $a=1$. 
The formal solution of this equation is the path-ordered exponent
\begin{equation}\label{eq:pexp}
  U(x,x_0)= P\exp\left[\epsilon\int\limits_{x_0}^{x} {\rm d}\xi M(\xi)\right]\;.
\end{equation}
This evolution operator can be readily expanded in $\epsilon$ with iterated integrals
as coefficients. Since we want to put boundary
conditions at $x=1$, we need to consider the limit $x_0\to 1$. 
Due to the presence of non-analytic terms of the form
$(1-x_0)^\epsilon$, this limit is not well defined when expanding in
$\epsilon$. This can be fixed by factoring out the non-analytic piece
\begin{equation}
  U(x,x_0)= \widetilde U(x,x_0)(1-x_0)^{-\epsilon M_1}\;,
\end{equation}
where $\widetilde U(x,x_0)$ has a finite limit for $x_0\to 1$. Therefore, we
can write down the general solution as
\begin{eqnarray}
  \mathbf{J}(x) &=& {\lim_{x_0\to1}U(x,x_0) \mathbf{C}_{x_0} }
  \nonumber\\
  &=& {\lim_{x_0\to1} U(x,x_0) (1-x_0)^{-\epsilon M_1} (1-x_0)^{\epsilon
      M_1} \mathbf{C}_{x_0} } 
  \nonumber\\
  &=& \widetilde U(x,1) \mathbf{C}\,, 
  \label{eq:general}
\end{eqnarray} 
where $\mathbf{C}_{x_0}$ and $\mathbf{C}$ are column vectors of constants (depending on
$\epsilon$). $\mathbf{C}_{x_0}$ depends in addition on $x_0$ whereas $\mathbf{C}$
does not. The ``reduced'' evolution operator $\widetilde U(x,1)$ can be
easily expanded in $\epsilon$ with the coefficients being harmonic
polylogarithms of $x$. The column of constants can be fixed by considering the
asymptotics of the canonical master integrals for $x\to 1$. In this limit
we have
\begin{equation}
  \mathbf{J}(x)\sim (1-x)^{\epsilon M_1} \mathbf{C}\,.
\end{equation}
Note, that we want to relate $\mathbf{C}$ to the coefficients of the asymptotic
expansion of the \emph{primary} master integrals, $\mathbf{j}(x)$, which are obtained from
the canonical ones via $\mathbf{j}(x) =
T(\epsilon,x) \mathbf{J}(x)$.\footnote{$T(\epsilon,x)$ is the transition
  matrix as introduced in Ref.~\cite{Lee:2014ioa} reducing the system to
  $\epsilon$ form.}
This might seem nontrivial since $T(\epsilon,x)$ (and
$T^{-1}(\epsilon,x)$) have
multiple poles for $x\to 1$. Therefore, we need to know several terms of
the asymptotic expansion of $ \mathbf{J}(x)$. We identically rewrite
Eq.~\eqref{eq:general} as
\begin{equation}
  \mathbf{J}(x) = (1-x)^{\epsilon M_1}{\overline{U}}(x,1)\mathbf{C}\,.
\end{equation}
The operator $\overline{U}(x,1)=(1-x)^{-\epsilon M_1}\widetilde
U(x,1)$ can be expanded in $1-x$ up to sufficiently high power which makes it
possible to connect the column of constants in $\mathbf{C}$ with the specific
coefficients of the asymptotic expansion of the primary master integrals at
$x=1$. 

We stress that the described method is extremely economic in the sense that
the overall number of asymptotic coefficients of the primary master
integrals (each being a function of $\epsilon$) to be fixed can be minimized
and is equal to the number of constants in $\mathbf{C}$. In addition, since
the integrals are all analytic in the point $x=1$, we set to zero all
coefficients in front of non-integer powers of $1-x$ in the generic
solution. This reduces even further the number of coefficients which we need
to calculate. We finally find that the boundary conditions are entirely fixed
by those entries of $\mathbf{j}(1)$ which are present among the 28 integrals
from Refs.~\cite{Baikov:2010hf,Lee:2011jt}.  Note that within our present
approach it is not necessary to calculate several expansion terms of the
primary masters near $x=1$, in contrast to Ref.~\cite{Lee:2016ixa}. The
analysis at $x=0$ is simplified in a similar way.

Repeating similar considerations for the point $x=0$, we finally connect the
coefficients of the asymptotic expansion of the primary master integrals for
$x\to 0$ with those for $x\to 1$. In particular, we extract the naive values
of the primary master integrals at $x=0$.

Following the procedure outlined in this Section we could compute all
master integrals contained in the families 7 and 786 up to weight 8.
Analytic expressions for the 24 non-planar master integrals
are given in the Appendix.


\section{\label{sec::res}$n_f^2$ results for form factors}

In this Section we discuss the results which we have obtained for the various
form factors using the techniques outlined in the previous Section and the
analytic results for the master integrals given in the Appendix.  We have used
a general QCD gauge parameter $\xi$ in the gluon propagator and have expanded
each Feynman amplitude up to the linear term.  We have checked that the
coefficient of $\xi$ vanishes for the bare form factors once all master
integrals are mapped to a minimal set.

  
\subsection{\label{sub::qff}Photon quark form factor}

For the $n_f^2$ term of the photon quark form factor
only the following three non-planar master integrals are needed:
\begin{eqnarray}
  G_{110110100111}^{\text{(7)}},
  G_{110110100112}^{\text{(7)}},
  G_{111101101110}^{\text{(786)}}
  \label{eq::npMIsnf2}
\end{eqnarray}
Analytic results are given in the Appendix.
We insert these results together with the ones for the planar master
integrals, expand in $\epsilon$ and renormalize $\alpha_s$.
After taking the logarithm we can compare to Eq.~(\ref{eq::logFx_gen})
and extract the cusp and collinear anomalous dimension.
We obtain
\begin{eqnarray}
  \gamma_{\rm cusp}^0 &=& 4
  \,,\nonumber\\
  \gamma_{\rm cusp}^1 &=&
  \left(\frac{268}{9}-\frac{4 \pi ^2}{3}\right) C_A-\frac{40 n_f}{9}
  \,,\nonumber\\
  \gamma_{\rm cusp}^2 &=&
  C_A^2 \left(\frac{88 \zeta _3}{3}+\frac{44 \pi ^4}{45}-\frac{536 \pi
      ^2}{27}+\frac{490}{3}\right)
  \nonumber\\&&\mbox{}
  +n_f \left[C_A \left(-\frac{112 \zeta
        _3}{3}+\frac{80 \pi ^2}{27}-\frac{836}{27}\right)+C_F \left(32 \zeta
      _3-\frac{110}{3}\right)\right]
  -\frac{16 n_f^2}{27}
  \,,\nonumber\\
  \gamma_{\rm cusp}^3 &=&
  \gamma _{\text{cusp}}^{3,n_f^0} + \gamma _{\text{cusp}}^{3,n_f^1}  n_f
  \nonumber\\&&\mbox{}
  +n_f^2 \Bigg[C_A \left(\frac{2240 \zeta _3}{27}-\frac{56 \pi
        ^4}{135}-\frac{304 \pi ^2}{243}+\frac{923}{81}\right)+C_F
    \left(-\frac{640 \zeta _3}{9}+\frac{16 \pi
        ^4}{45}+\frac{2392}{81}\right)\Bigg]
  \nonumber\\&&\mbox{}
  +\left(\frac{64 \zeta _3}{27}-\frac{32}{81}\right) n_f^3
  \,,
\label{eq::gamma_cusp}
\end{eqnarray}
and
\begin{eqnarray} 
  \gamma_{q}^0 &=& -3 C_F
  \,,\nonumber\\
  \gamma_{q}^1 &=&
   \left(26 \zeta _3-\frac{11 \pi ^2}{6}-\frac{961}{54}\right) C_A
    C_F+\left(\frac{65}{27}+\frac{\pi ^2}{3}\right) C_F n_f+\left(-24 \zeta
    _3+2 \pi ^2-\frac{3}{2}\right) C_F^2
  \,,\nonumber\\
  \gamma_{q}^2 &=&
   n_f \Bigg[\left(-\frac{964 \zeta _3}{27}+\frac{11 \pi ^4}{45}+\frac{1297
    \pi ^2}{243}-\frac{8659}{729}\right) C_A C_F+\left(\frac{256 \zeta
    _3}{9}-\frac{14 \pi ^4}{27}-\frac{13 \pi ^2}{9}
  \right.\nonumber\\&&\left.\mbox{}
    +\frac{2953}{54}\right)
    C_F^2\Bigg]
+\left(-\frac{8}{3} \pi ^2 \zeta _3-\frac{844 \zeta _3}{3}-120
    \zeta _5+\frac{247 \pi ^4}{135}+\frac{205 \pi ^2}{9}-\frac{151}{4}\right)
    C_A C_F^2
  \nonumber\\&&\mbox{}
+\left(-\frac{44}{9} \pi ^2 \zeta _3+\frac{3526 \zeta _3}{9}-136
    \zeta _5-\frac{83 \pi ^4}{90}-\frac{7163 \pi
    ^2}{486}-\frac{139345}{2916}\right) C_A^2 C_F
  \nonumber\\&&\mbox{}
+\left(-\frac{8 \zeta
    _3}{27}-\frac{10 \pi ^2}{27}+\frac{2417}{729}\right) C_F
    n_f^2+\left(\frac{16 \pi ^2 \zeta _3}{3}-68 \zeta _3+240 \zeta _5-\frac{8
    \pi ^4}{5}-3 \pi ^2
  \right.\nonumber\\&&\left.\mbox{}
-\frac{29}{2}\right) C_F^3
  \,,\nonumber\\
  \gamma_{q}^3 &=&
   n_f^2 \Bigg[\left(-\frac{64}{27} \pi ^2 \zeta _3-\frac{7436 \zeta
    _3}{243}+\frac{592 \zeta _5}{9}-\frac{19 \pi ^4}{135}-\frac{41579 \pi
    ^2}{8748}+\frac{97189}{34992}\right) C_A C_F
  \nonumber\\&&\mbox{}
+\left(\frac{56 \pi ^2 \zeta
    _3}{27}+\frac{2116 \zeta _3}{81}-\frac{520 \zeta _5}{9}+\frac{1004 \pi
    ^4}{1215}-\frac{493 \pi ^2}{81}-\frac{9965}{972}\right)
    C_F^2\Bigg]
  \nonumber\\&&\mbox{}
+\left(-\frac{712 \zeta _3}{243}-\frac{16 \pi
    ^4}{1215}-\frac{4 \pi ^2}{81}+\frac{18691}{6561}\right) C_F n_f^3+n_f
    \gamma _q^{3,n_f^1}+\gamma _q^{3,n_f^0}
  \,,
\label{eq::gamma_q}
\end{eqnarray}
where $\zeta_n$ is Riemann's zeta function evaluated at $n$.
The coefficients 
$\gamma _{\text{cusp}}^{3,n_f^0}$, $\gamma_{\text{cusp}}^{3,n_f^1}$,
$\gamma _{q}^{3,n_f^0}$ and $\gamma_{q}^{3,n_f^1}$ 
are only known in the large-$N_c$ limit~\cite{Henn:2016men,Lee:2016ixa}.
The one- to three-loop results for $\gamma _{\text{cusp}}$ and $\gamma_q$
can be found in
Refs.~\cite{Vogt:2000ci,Berger:2002sv,Moch:2004pa,Moch:2005tm,Baikov:2009bg,Becher:2009qa,Gehrmann:2010ue}
and the $n_f^3$ terms of $\gamma_{\text{cusp}}^3$ has been obtained in
Refs.~\cite{Gracey:1994nn,Beneke:1995pq}.
The $n_f^2$ term of $\gamma _{\text{cusp}}^{3}$ agrees with~\cite{Davies:2016jie}.
The terms in $\gamma _{q}^{3}$ which are beyond the large-$N_c$ limit are new.

For completeness we also present results for the finite part of $F_q$
which is conveniently done for the bare form factors since at each loop
order the $\mu$ dependence factorizes. In analogy to
Eq.~(\ref{eq::FFbare}) we write
\begin{eqnarray}
  \log(F_q) &=& 
  \sum_{n\ge1}
  \left(\frac{\alpha_s^0}{4\pi}\right)^n
  \left(\frac{\mu^2}{-q^2-i0} \right)^{n\epsilon}
  \log(F_q)|^{(n)}
  \,.
  \label{eq::logFq}
\end{eqnarray}
The $n_f^3$ and $n_f^2$ terms of $\log(F_q)|^{(4)}$ are given by
\begin{eqnarray}
\lefteqn{  \log(F_q)|^{(4)}_{n_f^2,n_f^3} =}\nonumber\\&&\mbox{} 
\frac{1}{\epsilon^5}  \Bigg[
\frac{1}{27} C_F n_f^3-\frac{11}{18} C_A C_F n_f^2
\Bigg]
%
%
+\frac{1}{\epsilon^4}  \Bigg[
n_f^2 \left(\left(\frac{\pi ^2}{18}-\frac{395}{54}\right) C_A C_F-\frac{2 C_F^2}{9}\right)+\frac{11}{27} C_F n_f^3
\Bigg]
\nonumber\\&&\mbox{}
+\frac{1}{\epsilon^3}  \Bigg[
n_f^2 \left(\left(\frac{19 \zeta _3}{3}-\frac{5 \pi
      ^2}{6}-\frac{75619}{1296}\right) C_A C_F+\left(-6 \zeta _3+\frac{2 \pi
      ^2}{3}-\frac{481}{108}\right) C_F^2\right)
\nonumber\\&&\mbox{}
+\left(\frac{254}{81}+\frac{5 \pi ^2}{81}\right) C_F n_f^3
\Bigg]
%
%
+\frac{1}{\epsilon^2}  \Bigg[
n_f^2 \left(\left(\frac{2170 \zeta _3}{27}+\frac{13 \pi ^4}{60}-\frac{1022 \pi
      ^2}{81}-\frac{2953141}{7776}\right) C_A C_F
\right.\nonumber\\&&\left.\mbox{}
+\left(-52 \zeta _3-\frac{7 \pi ^4}{90}+\frac{197 \pi ^2}{27}-\frac{14309}{324}\right) C_F^2\right)+\left(-\frac{82 \zeta _3}{81}+\frac{55 \pi ^2}{81}+\frac{29023}{1458}\right) C_F n_f^3
\Bigg]
\nonumber\\&&\mbox{}
+\frac{1}{\epsilon}  \Bigg[
n_f^2 \left(\left(-\frac{206}{27} \pi ^2 \zeta _3+\frac{61459 \zeta
      _3}{81}+\frac{784 \zeta _5}{9}+\frac{503 \pi ^4}{810}-\frac{424399 \pi
      ^2}{3888}-\frac{102630137}{46656}\right) C_A C_F
\right.\nonumber\\&&\left.\mbox{}
+\left(\frac{74 \pi ^2 \zeta _3}{9}-\frac{15121 \zeta _3}{27}+62 \zeta
  _5+\frac{31 \pi ^4}{135}+\frac{16993 \pi
    ^2}{324}-\frac{1308889}{3888}\right) C_F^2\right)
\nonumber\\&&\mbox{}
+\left(-\frac{902 \zeta _3}{81}+\frac{151 \pi ^4}{2430}+\frac{1270 \pi ^2}{243}+\frac{331889}{2916}\right) C_F n_f^3
\Bigg]
%
%
+ 
n_f^2 \left(\left(-\frac{1714 \zeta _3^2}{3}-\frac{218 \pi ^2 \zeta _3}{9}
\right.\right.\nonumber\\&&\left.\left.\mbox{}
+\frac{2897315 \zeta _3}{486}+\frac{150886 \zeta _5}{135}-\frac{709 \pi
  ^6}{17010}-\frac{1861 \pi ^4}{2430}-\frac{5825827 \pi
  ^2}{7776}-\frac{3325501813}{279936}\right) C_A C_F
\right.\nonumber\\&&\left.\mbox{}
+\left(\frac{2702 \zeta _3^2}{3}+\frac{1820 \pi ^2 \zeta _3}{27}-\frac{859249
    \zeta _3}{162}+\frac{1580 \zeta _5}{3}+\frac{17609 \pi
    ^6}{17010}+\frac{1141 \pi ^4}{1620}+\frac{76673 \pi ^2}{243}
\right.\right.\nonumber\\&&\left.\left.\mbox{}
-\frac{25891301}{11664}\right) C_F^2\right)+\left(-\frac{410}{243} \pi ^2
\zeta _3-\frac{20828 \zeta _3}{243}-\frac{2194 \zeta _5}{135}+\frac{1661 \pi
  ^4}{2430}+\frac{145115 \pi ^2}{4374}
\right.\nonumber\\&&\left.\mbox{}
+\frac{10739263}{17496}\right) C_F n_f^3
\,.
%
%
\end{eqnarray}
The four-loop term
in the large-$N_c$ limit can be extracted from Ref.~\cite{Lee:2016ixa};
all other terms are new.


\subsection{\label{sub::hbb}Higgs quark form factor}

The calculation of $F_b$ proceeds in close analogy to $F_q$.
It is interesting to note that about 20\% fewer integrals
are needed in the case of $F_b$. However, the complexity of the most
complicated integrals is the same and thus the CPU time
needed for the reduction to master integrals is comparable
for the two calculations.

After renormalizing $\alpha_s$ and $y$ and  taking the logarithm
we again compare to Eq.~(\ref{eq::logFx_gen})
and extract the cusp and collinear anomalous dimension. We obtain
the same results as in Eqs.~(\ref{eq::gamma_cusp}) and~(\ref{eq::gamma_q})
which constitutes a strong check for our calculation.

If we define $\log(F_b)$ expressed in terms of bare $\alpha_s$ and $y$ in
analogy to Eq.~(\ref{eq::logFq}) we get for the
$n_f^3$ and $n_f^2$ terms of $\log(F_b)|^{(4)}$
\begin{eqnarray}
\lefteqn{  \log(F_b)|^{(4)}_{n_f^2,n_f^3} =}\nonumber\\&&\mbox{}
\frac{1}{\epsilon^5}  \Bigg[
\frac{1}{27} C_F n_f^3-\frac{11}{18} C_A C_F n_f^2
\Bigg]
+\frac{1}{\epsilon^4}  \Bigg[
n_f^2 \left(\left(\frac{\pi ^2}{18}-\frac{197}{54}\right) C_A C_F-\frac{2 C_F^2}{9}\right)+\frac{5}{27} C_F n_f^3
\Bigg]
\nonumber\\&&\mbox{}
+\frac{1}{\epsilon^3}  \Bigg[
n_f^2 \left(\left(\frac{19 \zeta _3}{3}-\frac{5 \pi
      ^2}{6}-\frac{19819}{1296}\right) C_A C_F+\left(-6 \zeta _3+\frac{2 \pi
      ^2}{3}-\frac{283}{108}\right) C_F^2\right)
\nonumber\\&&\mbox{}
+\left(\frac{65}{81}+\frac{5 \pi ^2}{81}\right) C_F n_f^3
\Bigg]
+\frac{1}{\epsilon^2}  \Bigg[
n_f^2 \left(\left(\frac{1846 \zeta _3}{27}+\frac{13 \pi ^4}{60}-\frac{491 \pi
      ^2}{81}-\frac{417697}{7776}\right) C_A C_F
\right.\nonumber\\&&\left.\mbox{}
+\left(-64 \zeta _3-\frac{7 \pi ^4}{90}+\frac{119 \pi ^2}{27}-\frac{1244}{81}\right) C_F^2\right)+\left(-\frac{82 \zeta _3}{81}+\frac{25 \pi ^2}{81}+\frac{2537}{729}\right) C_F n_f^3
\Bigg]
\nonumber\\&&\mbox{}
+\frac{1}{\epsilon}  \Bigg[
n_f^2 \left(\left(-\frac{206}{27} \pi ^2 \zeta _3+\frac{35089 \zeta
      _3}{81}+\frac{784 \zeta _5}{9}+\frac{251 \pi ^4}{810}-\frac{121495 \pi
      ^2}{3888}-\frac{7185425}{46656}\right) C_A C_F
\right.\nonumber\\&&\left.\mbox{}
+\left(\frac{74 \pi ^2 \zeta _3}{9}-\frac{12079 \zeta _3}{27}+62 \zeta
  _5-\frac{38 \pi ^4}{135}+\frac{6871 \pi ^2}{324}-\frac{258229}{3888}\right)
C_F^2\right)+\left(-\frac{410 \zeta _3}{81}
\right.\nonumber\\&&\left.\mbox{}
+\frac{151 \pi ^4}{2430}+\frac{325 \pi ^2}{243}+\frac{11408}{729}\right) C_F n_f^3
\Bigg]
+ \Bigg[
n_f^2 \left(\left(-\frac{1714 \zeta _3^2}{3}+\frac{14 \pi ^2 \zeta
      _3}{9}+\frac{1209401 \zeta _3}{486}
\right.\right.\nonumber\\&&\left.\left.\mbox{}
+\frac{144946 \zeta _5}{135}-\frac{709 \pi ^6}{17010}+\frac{2129 \pi
  ^4}{2430}-\frac{1168375 \pi ^2}{7776}-\frac{73152481}{279936}\right) C_A
C_F+\left(\frac{2702 \zeta _3^2}{3}
\right.\right.\nonumber\\&&\left.\left.\mbox{}
+\frac{632 \pi ^2 \zeta
    _3}{27}-\frac{489655 \zeta _3}{162}-760 \zeta _5+\frac{17609 \pi
    ^6}{17010}-\frac{4693 \pi ^4}{1620}+\frac{94031 \pi
    ^2}{972}-\frac{2614421}{11664}\right) C_F^2\right)
\nonumber\\&&\mbox{}
+\left(-\frac{410}{243}
\pi ^2 \zeta _3-\frac{5330 \zeta _3}{243}-\frac{2194 \zeta _5}{135}+\frac{151
  \pi ^4}{486}+\frac{12685 \pi ^2}{2187}+\frac{159908}{2187}\right) C_F n_f^3 
\Bigg]\,.
\nonumber\\&&\mbox{}
\end{eqnarray}
To obtain this result the $\epsilon^2$ terms of the three-loop form factor are
needed. We refrain from listing them explicitly but present the expressions,
which are not yet available in the literature, in the ancillary file~\cite{progdata}.



\section{\label{sec::concl}Conclusions}

We have computed the complete $n_f^2$ contributions for the massless four-loop
fermionic form factors $F_q$ and $F_b$ and provide the corresponding cusp and
collinear anomalous dimensions.  This requires to consider two non-planar
integral families.  We systematically construct the solution applying
algorithmic procedures and obtain analytic results for all master integrals
contained in these families.  Although only three non-planar master integrals
are needed for the form factors we present analytic results for all 24
non-planar integrals, one of the main results of this paper. They constitute
important ingredients for future calculations, e.g., the $n_f^1$ and the
$n_f$-independent parts. Furthermore, we extend
the three-loop result for the Higgs fermion form factor to order
$\epsilon^2$. All analytic results can be downloaded from~\cite{progdata}.


\section*{\label{sec::ack}Acknowledgments}

This work is supported by the Deutsche Forschungsgemeinschaft through the
project ``Infrared and threshold effects in QCD''.  The work of A.S. and
V.S. is supported by RFBR, grant 17-02-00175A.


\begin{appendix}

\section*{Appendix: Explicit results for non-planar master integrals}

The 26 master integrals in family 7 (cf. Fig.~\ref{fig::diags_fam}) are
\begin{eqnarray}
&&
G_{000001111001}^{\text{(7)}}, G_{000000111111}^{\text{(7)}},
 G_{000001111110}^{\text{(7)}}
 , G_{000101111100}^{\text{(7)}}, G_{010000110111}^{\text{(7)}},
 G_{110000011011}^{\text{(7)}}
 , 
\nonumber\\&&
G_{000001111111}^{\text{(7)}}, G_{000101111110}^{\text{(7)}},
 G_{010000111111}^{\text{(7)}}
 , G_{010001111011}^{\text{(7)}}, G_{010001111012}^{\text{(7)}},
 G_{010110110101}^{\text{(7)}}
 , 
\nonumber\\&&
G_{011010110101}^{\text{(7)}}, G_{110001101011}^{\text{(7)}},
 G_{000110111111}^{\text{(7)}}
 , G_{001111111100}^{\text{(7)}}, G_{010010111111}^{\text{(7)}},
 G_{010110110111}^{\text{(7)}}
 , 
\nonumber\\&&
 G_{011010111101}^{\text{(7)}}, G_{011010111102}^{\text{(7)}},
 G_{011011111001}^{\text{(7)}},
  G_{110000111111}^{\text{(7)}}, G_{110001111011}^{\text{(7)}},
 G_{110011101011}^{\text{(7)}}
 , 
\nonumber\\&&
G_{110110100111}^{\text{(7)}}, G_{110110100112}^{\text{(7)}}
,
\end{eqnarray}
and the 40 master integrals in family 786 are chosen as
\begin{eqnarray}
&&
G_{001010011010}^{\text{(786)}}, G_{001001111010}^{\text{(786)}},
 G_{011010001110}^{\text{(786)}}
 , G_{100010011110}^{\text{(786)}}, G_{100101001110}^{\text{(786)}},
 G_{100101011010}^{\text{(786)}}
 , 
\nonumber\\&&
G_{110010010110}^{\text{(786)}}, G_{011001101110}^{\text{(786)}},
 G_{011001110110}^{\text{(786)}}
 , G_{011010011110}^{\text{(786)}}, G_{100001111110}^{\text{(786)}},
 G_{100101111010}^{\text{(786)}}
 , 
\nonumber\\&&
G_{100110011011}^{\text{(786)}}, G_{100110011110}^{\text{(786)}},
 G_{101001011110}^{\text{(786)}}
 , G_{101001011120}^{\text{(786)}}, G_{110001110110}^{\text{(786)}},
 G_{110110001110}^{\text{(786)}}
 , 
\nonumber\\&&
G_{111001001110}^{\text{(786)}}, G_{011001111110}^{\text{(786)}},
 G_{100101111110}^{\text{(786)}}
 , G_{100111011110}^{\text{(786)}}, G_{101001111110}^{\text{(786)}},
 G_{101011011110}^{\text{(786)}}
 , 
\nonumber\\&&
G_{101101111010}^{\text{(786)}}, G_{101110011011}^{\text{(786)}},
 G_{101111011010}^{\text{(786)}}
 , G_{110101101110}^{\text{(786)}}, G_{111001110110}^{\text{(786)}},
 G_{111010010111}^{\text{(786)}}
 ,
\nonumber\\&&
 G_{111010010112}^{\text{(786)}}, G_{111010011110}^{\text{(786)}},
 G_{111011010110}^{\text{(786)}}
 , G_{111110001110}^{\text{(786)}}, G_{101111011110}^{\text{(786)}},
 G_{111001111110}^{\text{(786)}}
 ,
\nonumber\\&&
 G_{111011011110}^{\text{(786)}}, G_{111011011120}^{\text{(786)}},
 G_{111101101110}^{\text{(786)}}
 , G_{111111001110}^{\text{(786)}}
.
\end{eqnarray}
The planar integrals have already been used in Ref.~\cite{Lee:2016ixa}
to obtain $F_q$ in the large-$N_c$ limit. Their analytic results
will be provided in Ref.~\cite{HSS}. Here, we concentrate on the non-planar
integrals. 

Family 7 has twelve non-planar master integrals. Five of them can be mapped
to family 786 using
\begin{eqnarray}
  G_{010110110111}^{\text{(7)}}&\to& G_{101001111110}^{\text{(786)}}
  \,,\nonumber\\
  G_{011011111001}^{\text{(7)}}&\to& G_{101110011011}^{\text{(786)}}
  \,,\nonumber\\
  G_{110001111011}^{\text{(7)}}&\to& G_{111010011110}^{\text{(786)}}
  \,,\nonumber\\
  G_{110011101011}^{\text{(7)}}&\to& G_{111110001110}^{\text{(786)}}
  \,,\nonumber\\
  G_{110110100111}^{\text{(7)}}&\to& G_{111010010111}^{\text{(786)}}
  \,.
\end{eqnarray}
The analytic results of the remaining seven integrals expanded up to weight eight are given by
\begin{eqnarray}
\lefteqn{G_{000110111111}^{\text{(7)}}=}
\nonumber\\&&
\frac{5 \zeta _5}{\epsilon}  
%
%
+ 
9 \zeta _3^2+55 \zeta _5+\frac{\pi ^6}{30}
%
%
+\epsilon  \Bigg[
99 \zeta _3^2+\frac{67 \pi ^4 \zeta _3}{90}-\frac{47 \pi ^2 \zeta _5}{3}+455 \zeta _5+\frac{591 \zeta _7}{2}+\frac{11 \pi ^6}{30}
\Bigg]
\nonumber\\&&\mbox{}
+\epsilon^2  \Bigg[
-\frac{76 s_{8 a}}{5}-5 \pi ^2 \zeta _3^2+819 \zeta _3^2-\frac{302 \zeta _5
  \zeta _3}{3}+\frac{737 \pi ^4 \zeta _3}{90}-\frac{517 \pi ^2 \zeta
  _5}{3}+3355 \zeta _5+\frac{6501 \zeta _7}{2}
\nonumber\\&&\mbox{}
+\frac{23689 \pi ^8}{378000}+\frac{91 \pi ^6}{30}
\Bigg]
\,,
%
%
\nonumber\\
\lefteqn{G_{001111111100}^{\text{(7)}}=}
\nonumber\\&&
\frac{1}{ 2\epsilon^4}
%
%
+\frac{7}{\epsilon^3}
%
%
+\frac{1}{\epsilon^2}  \Bigg[
63-\frac{\pi ^2}{2}
\Bigg]
%
%
+\frac{1}{\epsilon}  \Bigg[
-\frac{107 \zeta _3}{3}-7 \pi ^2+465
\Bigg]
%
%
+ 
-\frac{1498 \zeta _3}{3}-\frac{119 \pi ^4}{180}-63 \pi ^2
\nonumber\\&&\mbox{}
+3069
%
%
+\epsilon  \Bigg[
33 \pi ^2 \zeta _3-4506 \zeta _3-\frac{2437 \zeta _5}{5}-\frac{833 \pi ^4}{90}-465 \pi ^2+18873
\Bigg]
\nonumber\\&&\mbox{}
+\epsilon^2  \Bigg[
\frac{11593 \zeta _3^2}{9}+462 \pi ^2 \zeta _3-33446 \zeta _3-\frac{34118
  \zeta _5}{5}-\frac{6721 \pi ^6}{11340}-\frac{7519 \pi ^4}{90}-3069 \pi ^2
\nonumber\\&&\mbox{}
+110621
\Bigg]
%
%
+\epsilon^3  \Bigg[
\frac{162302 \zeta _3^2}{9}+\frac{12757 \pi ^4 \zeta _3}{270}+4170 \pi ^2
\zeta _3-222702 \zeta _3+\frac{6871 \pi ^2 \zeta _5}{15}
\nonumber\\&&\mbox{}
-\frac{307822 \zeta _5}{5}-\frac{46762 \zeta _7}{7}-\frac{6721 \pi ^6}{810}-\frac{55841 \pi ^4}{90}-18873 \pi ^2+626545
\Bigg]
\nonumber\\&&\mbox{}
+\epsilon^4  \Bigg[
-712 s_{8 a}-\frac{10169}{9} \pi ^2 \zeta _3^2+163134 \zeta _3^2+\frac{528478
  \zeta _5 \zeta _3}{15}+\frac{89299 \pi ^4 \zeta _3}{135}+30966 \pi ^2 \zeta
_3
\nonumber\\&&\mbox{}
-1386486 \zeta _3+\frac{96194 \pi ^2 \zeta _5}{15}-456778 \zeta _5-93524 \zeta
_7-\frac{43481 \pi ^8}{680400}-\frac{60439 \pi ^6}{810}-\frac{41349 \pi
  ^4}{10}
\nonumber\\&&\mbox{}
-110621 \pi ^2+3459765
\Bigg]
\,,
%
%
\nonumber\\
\lefteqn{G_{010010111111}^{\text{(7)}}=}
\nonumber\\&&
\frac{5 \zeta _5}{\epsilon}
%
%
+ 
7 \zeta _3^2-2 \pi ^2 \zeta _3+45 \zeta _5+\frac{71 \pi ^6}{2268}
%
\nonumber\\&&\mbox{}
+\epsilon  \Bigg[
47 \zeta _3^2-\frac{7 \pi ^4 \zeta _3}{18}-34 \pi ^2 \zeta _3+\frac{40 \pi ^2 \zeta _5}{3}+285 \zeta _5+\frac{907 \zeta _7}{4}+\frac{3023 \pi ^6}{11340}
\Bigg]
\nonumber\\&&\mbox{}
+\epsilon^2  \Bigg[
-\frac{582 s_{8 a}}{5}+\frac{53}{3} \pi ^2 \zeta _3^2+127 \zeta _3^2-\frac{722
  \zeta _5 \zeta _3}{3}-\frac{367 \pi ^4 \zeta _3}{90}-386 \pi ^2 \zeta _3+128
\pi ^2 \zeta _5
\nonumber\\&&\mbox{}
+1425 \zeta _5+\frac{9097 \zeta _7}{4}+\frac{341261 \pi ^8}{1701000}+\frac{2473 \pi ^6}{1620}
\Bigg]
\,,
%
\nonumber\\
\lefteqn{G_{011010111101}^{\text{(7)}}=}
\nonumber\\&&
\frac{3 \zeta _3}{2\epsilon^2}
%
%
+\frac{1}{\epsilon}  \Bigg[
\frac{39 \zeta _3}{2}+\frac{11 \pi ^4}{360}
\Bigg]
%
%
+ 
-\frac{5}{2} \pi ^2 \zeta _3+\frac{351 \zeta _3}{2}+\frac{3 \zeta _5}{2}+\frac{143 \pi ^4}{360}
%
\nonumber\\&&\mbox{}
+\epsilon  \Bigg[
-\frac{263 \zeta _3^2}{2}-\frac{57 \pi ^2 \zeta _3}{2}+\frac{2715 \zeta _3}{2}+\frac{79 \zeta _5}{2}-\frac{2519 \pi ^6}{22680}+\frac{143 \pi ^4}{40}
\Bigg]
\nonumber\\&&\mbox{}
+\epsilon^2  \Bigg[
-\frac{3299 \zeta _3^2}{2}-\frac{2491 \pi ^4 \zeta _3}{540}-\frac{433 \pi ^2
  \zeta _3}{2}+\frac{19383 \zeta _3}{2}-\frac{269 \pi ^2 \zeta
  _5}{6}+\frac{1111 \zeta _5}{2}
\nonumber\\&&\mbox{}
-\frac{2213 \zeta _7}{4}-\frac{29219 \pi ^6}{22680}+\frac{1991 \pi ^4}{72}
\Bigg]
%
%
+\epsilon^3  \Bigg[
316 s_{8 a}+\frac{755}{6} \pi ^2 \zeta _3^2-\frac{28491 \zeta
  _3^2}{2}-\frac{13531 \zeta _5 \zeta _3}{5}
\nonumber\\&&\mbox{}
-\frac{6329 \pi ^4 \zeta _3}{108}-\frac{2677 \pi ^2 \zeta _3}{2}+\frac{131859
  \zeta _3}{2}-\frac{3263 \pi ^2 \zeta _5}{6}+\frac{11955 \zeta
  _5}{2}-\frac{13445 \zeta _7}{2}-\frac{13939 \pi ^8}{24300}
\nonumber\\&&\mbox{}
-\frac{2811 \pi ^6}{280}+\frac{71071 \pi ^4}{360}
\Bigg]
\,,
%
%
\nonumber\\
\lefteqn{G_{011010111102}^{\text{(7)}}=}
\nonumber\\&&
\frac{1}{\epsilon^6}  \Bigg[
-\frac{1}{36}
\Bigg]
%
%
+\frac{1}{\epsilon^5}  \Bigg[
-\frac{1}{18}
\Bigg]
%
%
+\frac{1}{\epsilon^4}  \Bigg[
-\frac{1}{9}
\Bigg]
%
%
+\frac{1}{\epsilon^3}  \Bigg[
\frac{89 \zeta _3}{54}-\frac{2}{9}
\Bigg]
%
%
+\frac{1}{\epsilon^2}  \Bigg[
\frac{89 \zeta _3}{27}+\frac{7 \pi ^4}{120}-\frac{4}{9}
\Bigg]
\nonumber\\&&\mbox{}
+\frac{1}{\epsilon}  \Bigg[
\frac{7 \pi ^2 \zeta _3}{9}+\frac{178 \zeta _3}{27}+\frac{2117 \zeta _5}{90}+\frac{7 \pi ^4}{60}-\frac{8}{9}
\Bigg]
\nonumber\\&&\mbox{}
%
-\frac{9271 \zeta _3^2}{162}+\frac{14 \pi ^2 \zeta _3}{9}+\frac{356 \zeta _3}{27}+\frac{2117 \zeta _5}{45}+\frac{2573 \pi ^6}{34020}+\frac{7 \pi ^4}{30}-\frac{16}{9}
%
%
+\epsilon  \Bigg[
-\frac{9271 \zeta _3^2}{81}
\nonumber\\&&\mbox{}
-\frac{667 \pi ^4 \zeta _3}{180}+\frac{28 \pi ^2 \zeta _3}{9}+\frac{712 \zeta _3}{27}+9 \pi ^2 \zeta _5+\frac{4234 \zeta _5}{45}+\frac{21151 \zeta _7}{252}+\frac{2573 \pi ^6}{17010}+\frac{7 \pi ^4}{15}-\frac{32}{9}
\Bigg]
\nonumber\\&&\mbox{}
+\epsilon^2  \Bigg[
\frac{848 s_{8 a}}{5}-\frac{1528}{27} \pi ^2 \zeta _3^2-\frac{18542 \zeta
  _3^2}{81}-\frac{230803 \zeta _5 \zeta _3}{135}-\frac{667 \pi ^4 \zeta
  _3}{90}+\frac{344 \pi ^2 \zeta _3}{9}+\frac{1424 \zeta _3}{27}
\nonumber\\&&\mbox{}
+18 \pi ^2 \zeta _5+\frac{15668 \zeta _5}{45}+\frac{21151 \zeta _7}{126}-\frac{500903 \pi ^8}{3402000}+\frac{2573 \pi ^6}{8505}+\frac{14 \pi ^4}{15}-\frac{64}{9}
\Bigg]
\,,
%
%
%
%
\nonumber\\
\lefteqn{G_{110000111111}^{\text{(7)}}=}
\nonumber\\&&
+\frac{1}{\epsilon^4}  \Bigg[
\frac{1}{12}
\Bigg]
%
%
+\frac{1}{\epsilon^3}  
%
%
+\frac{1}{\epsilon^2}  \Bigg[
\frac{91}{12}-\frac{\pi ^2}{36}
\Bigg]
%
%
+\frac{1}{\epsilon}  \Bigg[
-\frac{37 \zeta _3}{9}-\frac{\pi ^2}{3}+\frac{93}{2}
\Bigg]
%
%
+ 
\frac{\pi ^2 \zeta _3}{3}-48 \zeta _3+\frac{5 \zeta _5}{3}
\nonumber\\&&\mbox{}
-\frac{37 \pi ^4}{540}-\frac{95 \pi ^2}{36}+\frac{3025}{12}
%
%
+\epsilon  \Bigg[
\frac{5 \zeta _3^2}{3}+\frac{115 \pi ^2 \zeta _3}{27}-\frac{3131 \zeta
  _3}{9}-\frac{2038 \zeta _5}{45}+\frac{88 \pi ^6}{8505}-\frac{193 \pi
  ^4}{270}
\nonumber\\&&\mbox{}
-\frac{955 \pi ^2}{54}+1262
\Bigg]
%
%
+\epsilon^2  \Bigg[
\frac{2543 \zeta _3^2}{27}+\frac{71 \pi ^4 \zeta _3}{270}+\frac{1039 \pi ^2
  \zeta _3}{27}-\frac{53390 \zeta _3}{27}-\frac{\pi ^2 \zeta
  _5}{9}-\frac{79163 \zeta _5}{135}
\nonumber\\&&\mbox{}
+\frac{407 \zeta _7}{12}-\frac{799 \pi ^6}{14580}-\frac{7261 \pi ^4}{1620}-\frac{35449 \pi ^2}{324}+\frac{71767}{12}
\Bigg]
%
%
+\epsilon^3  \Bigg[
-\frac{346 s_{8 a}}{15}-\frac{94}{3} \pi ^2 \zeta _3^2
\nonumber\\&&\mbox{}
+\frac{25763 \zeta _3^2}{27}-\frac{1694 \zeta _5 \zeta _3}{9}+\frac{1073 \pi
  ^4 \zeta _3}{270}+\frac{26051 \pi ^2 \zeta _3}{81}-\frac{761281 \zeta
  _3}{81}+\frac{383 \pi ^2 \zeta _5}{135}-\frac{1657792 \zeta _5}{405}
\nonumber\\&&\mbox{}
-\frac{278843 \zeta _7}{252}-\frac{461 \pi ^8}{70875}-\frac{14227 \pi ^6}{15309}-\frac{24929 \pi ^4}{1215}-\frac{157496 \pi ^2}{243}+\frac{54491}{2}
\Bigg]
\,,
%
%
\nonumber\\
\lefteqn{G_{110110100112}^{\text{(7)}}=}
\nonumber\\&&
+\frac{1}{\epsilon^4}  \Bigg[
-\frac{\pi ^2}{36}
\Bigg]
%
%
+\frac{1}{\epsilon^3}  \Bigg[
-\frac{\zeta _3}{3}-\frac{\pi ^2}{18}
\Bigg]
%
%
+\frac{1}{\epsilon^2}  \Bigg[
-\frac{2 \zeta _3}{3}-\frac{\pi ^4}{30}-\frac{\pi ^2}{9}
\Bigg]
\nonumber\\&&\mbox{}
+\frac{1}{\epsilon}  \Bigg[
\frac{37 \pi ^2 \zeta _3}{27}-\frac{4 \zeta _3}{3}+6 \zeta _5-\frac{\pi ^4}{15}-\frac{2 \pi ^2}{9}
\Bigg]
%
%
+ 
\frac{361 \zeta _3^2}{9}+\frac{74 \pi ^2 \zeta _3}{27}-\frac{8 \zeta _3}{3}+12
\zeta _5
+\frac{71 \pi ^6}{1890}
\nonumber\\&&\mbox{}
-\frac{2 \pi ^4}{15}-\frac{4 \pi ^2}{9}
%
%
+\epsilon  \Bigg[
\frac{722 \zeta _3^2}{9}+\frac{44 \pi ^4 \zeta _3}{15}+\frac{148 \pi ^2 \zeta
  _3}{27}-\frac{16 \zeta _3}{3}+\frac{886 \pi ^2 \zeta _5}{45}+24 \zeta
_5+\frac{3925 \zeta _7}{8}
\nonumber\\&&\mbox{}
+\frac{71 \pi ^6}{945}-\frac{4 \pi ^4}{15}-\frac{8 \pi ^2}{9}
\Bigg]
%
%
+\epsilon^2  \Bigg[
-\frac{869 s_{8 a}}{5}-\frac{2738}{81} \pi ^2 \zeta _3^2+\frac{1444 \zeta
  _3^2}{9}+\frac{11224 \zeta _5 \zeta _3}{15}+\frac{88 \pi ^4 \zeta _3}{15}
\nonumber\\&&\mbox{}
+\frac{296 \pi ^2 \zeta _3}{27}-\frac{32 \zeta _3}{3}+\frac{1772 \pi ^2 \zeta
  _5}{45}+48 \zeta _5+\frac{3925 \zeta _7}{4}+\frac{365909 \pi
  ^8}{972000}+\frac{142 \pi ^6}{945}-\frac{8 \pi ^4}{15}
\nonumber\\&&\mbox{}
-\frac{16 \pi ^2}{9}
\Bigg]\,,
%
%
\end{eqnarray}
where
\begin{eqnarray} 
  s_{8a} &=& \zeta_8 + \zeta_{5,3} \approx 1.0417850291827918834\,.
\end{eqnarray}
$\zeta_{m_{1},\dots,m_{k}}$ are multiple zeta values given by
\begin{eqnarray}
  \zeta_{m_{1},\dots,m_{k}} &=&
  \sum\limits _{i_{1}=1}^{\infty}\sum\limits
  _{i_{2}=1}^{i_{1}-1}\dots\sum\limits _{i_{k}=1}^{i_{k-1}-1}\prod\limits
  _{j=1}^{k}\frac{\mbox{sgn}(m_{j})^{i_{j}}}{i_{j}^{|m_{j}|}}
  \,. 
\end{eqnarray}

Family 786 has 17 non-planar master integrals. Their analytic results read
\begin{eqnarray}
\lefteqn{G_{101001111110}^{\text{(786)}}=}
\nonumber\\&&
+\frac{1}{\epsilon^2}  \Bigg[
\frac{3 \zeta _3}{2}
\Bigg]
%
%
+\frac{1}{\epsilon}  \Bigg[
\frac{39 \zeta _3}{2}+\frac{11 \pi ^4}{360}
\Bigg]
%
%
+ \Bigg[
-\frac{1}{2} \pi ^2 \zeta _3+\frac{351 \zeta _3}{2}+\frac{23 \zeta _5}{2}+\frac{143 \pi ^4}{360}
\Bigg]
\nonumber\\&&\mbox{}
+\epsilon  \Bigg[
-\frac{211 \zeta _3^2}{2}-\frac{13 \pi ^2 \zeta _3}{2}+\frac{2715 \zeta _3}{2}+\frac{299 \zeta _5}{2}-\frac{121 \pi ^6}{3240}+\frac{143 \pi ^4}{40}
\Bigg]
\nonumber\\&&\mbox{}
+\epsilon^2  \Bigg[
-\frac{2743 \zeta _3^2}{2}-\frac{2257 \pi ^4 \zeta _3}{540}-\frac{117 \pi ^2
  \zeta _3}{2}+\frac{19383 \zeta _3}{2}+\frac{28 \pi ^2 \zeta
  _5}{3}+\frac{2691 \zeta _5}{2}-\frac{1925 \zeta _7}{4}
\nonumber\\&&\mbox{}
-\frac{1573 \pi ^6}{3240}+\frac{1991 \pi ^4}{72}
\Bigg]
%
%
+\epsilon^3  \Bigg[
\frac{1603 s_{8 a}}{5}+\frac{98}{3} \pi ^2 \zeta _3^2-\frac{24687 \zeta
  _3^2}{2}-\frac{28928 \zeta _5 \zeta _3}{15}-\frac{29341 \pi ^4 \zeta
  _3}{540}
\nonumber\\&&\mbox{}
-\frac{905 \pi ^2 \zeta _3}{2}+\frac{131859 \zeta _3}{2}+\frac{364 \pi ^2
  \zeta _5}{3}+\frac{20815 \zeta _5}{2}-\frac{25025 \zeta _7}{4}-\frac{907097
  \pi ^8}{2268000}-\frac{1573 \pi ^6}{360}
\nonumber\\&&\mbox{}
+\frac{71071 \pi ^4}{360}
\Bigg]\,,
%
%
\nonumber\\
\lefteqn{G_{101011011110}^{\text{(786)}}=}
\nonumber\\&&
+\frac{1}{\epsilon^2}  \Bigg[
-\frac{3 \zeta _3}{2}
\Bigg]
%
%
+\frac{1}{\epsilon}  \Bigg[
-\frac{33 \zeta _3}{2}-\frac{\pi ^4}{40}
\Bigg]
%
%
+ \Bigg[
-\frac{5}{6} \pi ^2 \zeta _3-\frac{201 \zeta _3}{2}-\frac{71 \zeta _5}{2}-\frac{91 \pi ^4}{360}
\Bigg]
\nonumber\\&&\mbox{}
+\epsilon  \Bigg[
\frac{5 \zeta _3^2}{2}-\frac{33 \pi ^2 \zeta _3}{2}-\frac{501 \zeta _3}{2}-\frac{913 \zeta _5}{2}-\frac{3673 \pi ^6}{22680}-\frac{29 \pi ^4}{24}
\Bigg]
\nonumber\\&&\mbox{}
+\epsilon^2  \Bigg[
-\frac{793 \zeta _3^2}{2}-\frac{283 \pi ^4 \zeta _3}{180}-\frac{1259 \pi ^2
  \zeta _3}{6}+\frac{5991 \zeta _3}{2}+\frac{79 \pi ^2 \zeta _5}{2}-\frac{7529
  \zeta _5}{2}-\frac{6675 \zeta _7}{4}
\nonumber\\&&\mbox{}
-\frac{18701 \pi ^6}{7560}+\frac{713 \pi ^4}{360}
\Bigg]
%
%
+\epsilon^3  \Bigg[
\frac{4632 s_{8 a}}{5}+\frac{2339}{18} \pi ^2 \zeta _3^2-\frac{17473 \zeta
  _3^2}{2}+\frac{26513 \zeta _5 \zeta _3}{15}-\frac{4421 \pi ^4 \zeta _3}{108}
\nonumber\\&&\mbox{}
-\frac{4341 \pi ^2 \zeta _3}{2}+\frac{118371 \zeta _3}{2}+\frac{1107 \pi ^2
  \zeta _5}{2}-\frac{48421 \zeta _5}{2}-\frac{51301 \zeta _7}{2}-\frac{1726121
  \pi ^8}{1701000}
\nonumber\\&&\mbox{}
-\frac{575791 \pi ^6}{22680}+\frac{13823 \pi ^4}{120}
\Bigg]\,,
%
%
\nonumber\\
\lefteqn{G_{101101111010}^{\text{(786)}}=}
\nonumber\\&&
\frac{1}{4\epsilon^4}
%
%
+\frac{1}{\epsilon^3}  \Bigg[
\frac{13}{4}-\frac{\pi ^2}{12}
\Bigg]
%
%
+\frac{1}{\epsilon^2}  \Bigg[
-\frac{5 \zeta _3}{2}-\frac{5 \pi ^2}{6}+\frac{109}{4}
\Bigg]
%
%
+\frac{1}{\epsilon}  \Bigg[
-\frac{112 \zeta _3}{3}-\frac{19 \pi ^4}{144}-\frac{35 \pi ^2}{6}
\nonumber\\&&\mbox{}
+\frac{753}{4}
\Bigg]
%
%
+ 
\frac{157 \pi ^2 \zeta _3}{36}-\frac{1024 \zeta _3}{3}-\frac{125 \zeta _5}{4}-\frac{1043 \pi ^4}{720}-\frac{71 \pi ^2}{2}+\frac{4677}{4}
%
\nonumber\\&&\mbox{}
+\epsilon  \Bigg[
\frac{953 \zeta _3^2}{6}+\frac{803 \pi ^2 \zeta _3}{18}-\frac{7570 \zeta _3}{3}-\frac{2471 \zeta _5}{5}-\frac{893 \pi ^6}{18144}-\frac{883 \pi ^4}{80}-\frac{403 \pi ^2}{2}+\frac{27225}{4}
\Bigg]
\nonumber\\&&\mbox{}
+\epsilon^2  \Bigg[
\frac{16979 \zeta _3^2}{9}+\frac{20909 \pi ^4 \zeta _3}{2160}+\frac{2896 \pi
  ^2 \zeta _3}{9}-16670 \zeta _3+\frac{6563 \pi ^2 \zeta _5}{120}-\frac{46481
  \zeta _5}{10}+\frac{3805 \zeta _7}{32}
\nonumber\\&&\mbox{}
-\frac{75931 \pi ^6}{90720}-\frac{52279 \pi ^4}{720}-\frac{2199 \pi ^2}{2}+\frac{151933}{4}
\Bigg]
%
%
+\epsilon^3  \Bigg[
-299 s_{8 a}-\frac{3638}{27} \pi ^2 \zeta _3^2
\nonumber\\&&\mbox{}
+\frac{138134 \zeta _3^2}{9}+\frac{26623 \zeta _5 \zeta _3}{6}+\frac{22397 \pi
  ^4 \zeta _3}{216}+\frac{6100 \pi ^2 \zeta _3}{3}-103214 \zeta _3+\frac{6773
  \pi ^2 \zeta _5}{12}
\nonumber\\&&\mbox{}
-\frac{349893 \zeta _5}{10}-\frac{108151 \zeta _7}{56}+\frac{1637173 \pi
  ^8}{3628800}-\frac{241853 \pi ^6}{30240}-\frac{319363 \pi
  ^4}{720}-\frac{35129 \pi ^2}{6}
\nonumber\\&&\mbox{}
+\frac{823921}{4}
\Bigg]\,,
%
%
\nonumber\\
\lefteqn{G_{101110011011}^{\text{(786)}}=}
\nonumber\\&&
-\frac{1}{4\epsilon^4} 
%
%
-\frac{13}{4\epsilon^3} 
%
%
+\frac{1}{\epsilon^2}  \Bigg[
\frac{\pi ^2}{12}-\frac{109}{4}
\Bigg]
%
%
+\frac{1}{\epsilon}  \Bigg[
\frac{40 \zeta _3}{3}+\frac{13 \pi ^2}{12}-\frac{753}{4}
\Bigg]
%
%
+ 
\frac{538 \zeta _3}{3}+\frac{53 \pi ^4}{180}+\frac{109 \pi ^2}{12}
\nonumber\\&&\mbox{}
-\frac{4677}{4}
%
%
+\epsilon  \Bigg[
-\frac{46}{9} \pi ^2 \zeta _3+\frac{4702 \zeta _3}{3}+\frac{846 \zeta _5}{5}+\frac{79 \pi ^4}{20}+\frac{251 \pi ^2}{4}-\frac{27225}{4}
\Bigg]
\nonumber\\&&\mbox{}
+\epsilon^2  \Bigg[
-\frac{4055 \zeta _3^2}{9}-\frac{625 \pi ^2 \zeta _3}{9}+11378 \zeta
_3+\frac{11428 \zeta _5}{5}+\frac{1345 \pi ^6}{4536}+\frac{413 \pi
  ^4}{12}+\frac{1559 \pi ^2}{4}
\nonumber\\&&\mbox{}
-\frac{151933}{4}
\Bigg]
+\epsilon^3  \Bigg[
-\frac{56162 \zeta _3^2}{9}-\frac{2107 \pi ^4 \zeta _3}{108}-\frac{5527 \pi ^2
  \zeta _3}{9}+74906 \zeta _3-\frac{706 \pi ^2 \zeta _5}{15}
\nonumber\\&&\mbox{}
+\frac{100384 \zeta _5}{5}+\frac{89139 \zeta _7}{56}+\frac{1963 \pi ^6}{504}+\frac{8963 \pi ^4}{36}+\frac{9075 \pi ^2}{4}-\frac{823921}{4}
\Bigg]
\nonumber\\&&\mbox{}
+\epsilon^4  \Bigg[
\frac{3308 s_{8 a}}{5}+\frac{10597}{54} \pi ^2 \zeta _3^2-\frac{507488 \zeta
  _3^2}{9}-11699 \zeta _5 \zeta _3-\frac{808 \pi ^4 \zeta _3}{3}-\frac{13553
  \pi ^2 \zeta _3}{3}
\nonumber\\&&\mbox{}
+466338 \zeta _3-\frac{9583 \pi ^2 \zeta _5}{15}+\frac{732928 \zeta
  _5}{5}+\frac{1154873 \zeta _7}{56}-\frac{1191469 \pi
  ^8}{4536000}+\frac{50021 \pi ^6}{1512}
\nonumber\\&&\mbox{}
+\frac{97961 \pi ^4}{60}+\frac{151933 \pi ^2}{12}-\frac{4378933}{4}
\Bigg]\,,
%
%
\nonumber\\
\lefteqn{G_{101111011010}^{\text{(786)}}=}
\nonumber\\&&
\frac{1}{4\epsilon^4}
%
%
+\frac{15}{4\epsilon^3}
%
%
+\frac{1}{\epsilon^2}  \Bigg[
-3 \zeta _3-\frac{\pi ^2}{4}+\frac{143}{4}
\Bigg]
%
%
+\frac{1}{\epsilon}  \Bigg[
-\frac{293 \zeta _3}{6}-\frac{23 \pi ^4}{360}-\frac{11 \pi ^2}{3}+\frac{1107}{4}
\Bigg]
\nonumber\\&&\mbox{}
+ \Bigg[
\frac{17 \pi ^2 \zeta _3}{6}-\frac{2017 \zeta _3}{4}-\frac{77 \zeta _5}{2}-\frac{779 \pi ^4}{720}-\frac{821 \pi ^2}{24}+\frac{7599}{4}
\Bigg]
%
%
+\epsilon  \Bigg[
246 \zeta _3^2
\nonumber\\&&\mbox{}
+\frac{187 \pi ^2 \zeta _3}{4}-\frac{102037 \zeta _3}{24}-\frac{12809 \zeta _5}{20}+\frac{487 \pi ^6}{7560}-\frac{16277 \pi ^4}{1440}-\frac{4151 \pi ^2}{16}+\frac{48267}{4}
\Bigg]
\nonumber\\&&\mbox{}
+\epsilon^2  \Bigg[
\frac{29987 \zeta _3^2}{9}+\frac{290 \pi ^4 \zeta _3}{27}+\frac{34609 \pi ^2
  \zeta _3}{72}-\frac{513681 \zeta _3}{16}+\frac{52 \pi ^2 \zeta
  _5}{3}-\frac{53407 \zeta _5}{8}
\nonumber\\&&\mbox{}
+\frac{1771 \zeta _7}{2}+\frac{7391 \pi ^6}{15120}-\frac{54919 \pi ^4}{576}-\frac{55901 \pi ^2}{32}+\frac{290551}{4}
\Bigg]
%
%
+\epsilon^3  \Bigg[
-1104 s_{8 a}-\frac{2014}{9} \pi ^2 \zeta _3^2
\nonumber\\&&\mbox{}
+\frac{62409 \zeta _3^2}{2}+\frac{96836 \zeta _5 \zeta _3}{15}+\frac{19859 \pi
  ^4 \zeta _3}{135}+\frac{577855 \pi ^2 \zeta _3}{144}-\frac{7274999 \zeta
  _3}{32}+\frac{5638 \pi ^2 \zeta _5}{15}
\nonumber\\&&\mbox{}
-\frac{4519813 \zeta _5}{80}+\frac{231549 \zeta _7}{28}+\frac{46621 \pi
  ^8}{37800}+\frac{70573 \pi ^6}{30240}-\frac{458701 \pi ^4}{640}-\frac{697507
  \pi ^2}{64}
\nonumber\\&&\mbox{}
+\frac{1682259}{4}
\Bigg]\,,
%
%
\nonumber\\
\lefteqn{G_{111001110110}^{\text{(786)}}=}
\nonumber\\&&
\frac{1}{12\epsilon^4}
%
%
+\frac{13}{12\epsilon^3} 
%
%
+\frac{1}{\epsilon^2}  \Bigg[
\frac{109}{12}-\frac{\pi ^2}{36}
\Bigg]
%
%
+\frac{1}{\epsilon}  \Bigg[
-\frac{28 \zeta _3}{9}-\frac{13 \pi ^2}{36}+\frac{251}{4}
\Bigg]
+ 
-\frac{418 \zeta _3}{9}
\nonumber\\&&\mbox{}
-\frac{29 \pi ^4}{270}-\frac{109 \pi ^2}{36}+\frac{1559}{4}
%
%
+\epsilon  \Bigg[
\frac{55 \pi ^2 \zeta _3}{27}-\frac{4078 \zeta _3}{9}-\frac{16 \zeta _5}{15}-\frac{41 \pi ^4}{27}-\frac{251 \pi ^2}{12}+\frac{9075}{4}
\Bigg]
\nonumber\\&&\mbox{}
+\epsilon^2  \Bigg[
\frac{3593 \zeta _3^2}{27}+\frac{742 \pi ^2 \zeta _3}{27}-\frac{11042 \zeta
  _3}{3}-\frac{1873 \zeta _5}{15}+\frac{293 \pi ^6}{7560}-\frac{1894 \pi
  ^4}{135}-\frac{1559 \pi ^2}{12}
\nonumber\\&&\mbox{}
+\frac{151933}{12}
\Bigg]
%
%
+\epsilon^3  \Bigg[
\frac{56888 \zeta _3^2}{27}+\frac{2849 \pi ^4 \zeta _3}{324}+\frac{6508 \pi ^2
  \zeta _3}{27}-\frac{81290 \zeta _3}{3}+\frac{1067 \pi ^2 \zeta
  _5}{90}-\frac{33379 \zeta _5}{15}
\nonumber\\&&\mbox{}
+\frac{107831 \zeta _7}{84}+\frac{2927 \pi ^6}{7560}-\frac{1622 \pi ^4}{15}-\frac{3025 \pi ^2}{4}+\frac{823921}{12}
\Bigg]
%
%
+\epsilon^4  \Bigg[
-\frac{2556 s_{8 a}}{5}-\frac{6131}{81} \pi ^2 \zeta _3^2
\nonumber\\&&\mbox{}
+\frac{585038 \zeta _3^2}{27}+\frac{101287 \zeta _5 \zeta _3}{45}+\frac{51769
  \pi ^4 \zeta _3}{405}+\frac{15812 \pi ^2 \zeta _3}{9}-188038 \zeta
_3+\frac{7408 \pi ^2 \zeta _5}{45}
\nonumber\\&&\mbox{}
-\frac{127781 \zeta _5}{5}+\frac{2737267 \zeta _7}{168}+\frac{1794577 \pi
  ^8}{1944000}+\frac{15179 \pi ^6}{7560}-\frac{34106 \pi ^4}{45}-\frac{151933
  \pi ^2}{36}
\nonumber\\&&\mbox{}
+\frac{4378933}{12}
\Bigg]\,,
%
%
\nonumber\\
\lefteqn{G_{111010011110}^{\text{(786)}}=}
\nonumber\\&&
\frac{1}{24\epsilon^4}  
%
%
+\frac{7}{12\epsilon^3}
%
%
+\frac{1}{\epsilon^2}  \Bigg[
\frac{125}{24}-\frac{\pi ^2}{72}
\Bigg]
%
%
+\frac{1}{\epsilon}  \Bigg[
-\frac{23 \zeta _3}{9}-\frac{7 \pi ^2}{36}+\frac{455}{12}
\Bigg]
\nonumber\\&&\mbox{}
%
-\frac{349 \zeta _3}{9}-\frac{79 \pi ^4}{1080}-\frac{125 \pi ^2}{72}+\frac{1967}{8}
%
%
+\epsilon  \Bigg[
\frac{50 \pi ^2 \zeta _3}{27}-\frac{3415 \zeta _3}{9}-\frac{158 \zeta
  _5}{15}-\frac{293 \pi ^4}{270}
\nonumber\\&&\mbox{}
-\frac{455 \pi ^2}{36}+\frac{5929}{4}
\Bigg]
%
%
+\epsilon^2  \Bigg[
\frac{3493 \zeta _3^2}{27}+\frac{673 \pi ^2 \zeta _3}{27}-\frac{27545 \zeta
  _3}{9}-\frac{3007 \zeta _5}{15}+\frac{83 \pi ^6}{2835}
\nonumber\\&&\mbox{}
-\frac{2239 \pi ^4}{216}-\frac{1967 \pi ^2}{24}+\frac{204205}{24}
\Bigg]
%
%
+\epsilon^3  \Bigg[
\frac{53789 \zeta _3^2}{27}+\frac{10697 \pi ^4 \zeta _3}{1620}+\frac{5629 \pi
  ^2 \zeta _3}{27}-\frac{66661 \zeta _3}{3}
\nonumber\\&&\mbox{}
+\frac{1538 \pi ^2 \zeta _5}{45}-\frac{7130 \zeta _5}{3}+\frac{98507 \zeta _7}{168}+\frac{205 \pi ^6}{648}-\frac{4403 \pi ^4}{54}-\frac{5929 \pi ^2}{12}+\frac{566335}{12}
\Bigg]
\nonumber\\&&\mbox{}
+\epsilon^4  \Bigg[
-\frac{1677 s_{8 a}}{5}-\frac{9011}{162} \pi ^2 \zeta _3^2+\frac{534284 \zeta
  _3^2}{27}+\frac{112502 \zeta _5 \zeta _3}{45}+\frac{32201 \pi ^4 \zeta
  _3}{324}+\frac{73315 \pi ^2 \zeta _3}{54}
\nonumber\\&&\mbox{}
-\frac{454363 \zeta _3}{3}+\frac{84103 \pi ^2 \zeta _5}{180}-\frac{67891 \zeta
  _5}{3}+\frac{191359 \zeta _7}{24}+\frac{1062491 \pi ^8}{1701000}+\frac{79579
  \pi ^6}{45360}
\nonumber\\&&\mbox{}
-\frac{69251 \pi ^4}{120}-\frac{204205 \pi ^2}{72}+\frac{6129101}{24}
\Bigg]\,,
%
%
\nonumber\\
\lefteqn{G_{111011010110}^{\text{(786)}}=}
\nonumber\\&&
-\frac{1}{12\epsilon^4}
%
%
-\frac{11}{12\epsilon^3} 
%
%
+\frac{1}{\epsilon^2}  \Bigg[
-\frac{71}{12}-\frac{5 \pi ^2}{36}
\Bigg]
%
%
+\frac{1}{\epsilon}  \Bigg[
\frac{28 \zeta _3}{9}-\frac{67 \pi ^2}{36}-\frac{319}{12}
\Bigg]
\nonumber\\&&\mbox{}
+
\frac{463 \zeta _3}{18}-\frac{31 \pi ^4}{270}-\frac{287 \pi ^2}{18}-\frac{269}{4}
%
%
+\epsilon  \Bigg[
\frac{731 \pi ^2 \zeta _3}{108}+\frac{1841 \zeta _3}{36}+\frac{4339 \zeta
  _5}{60}-\frac{1997 \pi ^4}{1080}
\nonumber\\&&\mbox{}
-\frac{8047 \pi ^2}{72}+\frac{835}{4}
\Bigg]
%
%
+\epsilon^2  \Bigg[
\frac{4609 \zeta _3^2}{108}+\frac{20273 \pi ^2 \zeta _3}{216}-\frac{78925
  \zeta _3}{72}+\frac{88993 \zeta _5}{120}+\frac{2459 \pi ^6}{12960}
\nonumber\\&&\mbox{}
-\frac{41221 \pi ^4}{2160}-\frac{33637 \pi ^2}{48}+\frac{54833}{12}
\Bigg]
\nonumber\\&&\mbox{}
+\epsilon^3  \Bigg[
\frac{287959 \zeta _3^2}{216}+\frac{34867 \pi ^4 \zeta _3}{3240}+\frac{363883
  \pi ^2 \zeta _3}{432}-\frac{865033 \zeta _3}{48}+\frac{1247 \pi ^2 \zeta
  _5}{18}+\frac{939131 \zeta _5}{240}
\nonumber\\&&\mbox{}
+\frac{1875815 \zeta _7}{672}+\frac{404363 \pi ^6}{181440}-\frac{697471 \pi ^4}{4320}-\frac{393983 \pi ^2}{96}+\frac{516097}{12}
\Bigg]
%
%
+\epsilon^4  \Bigg[
-\frac{9407 s_{8 a}}{10}
\nonumber\\&&\mbox{}-\frac{69871}{324} \pi ^2 \zeta _3^2+\frac{8620349 \zeta
  _3^2}{432}+\frac{6038 \zeta _5 \zeta _3}{45}+\frac{1137997 \pi ^4 \zeta
  _3}{6480}+\frac{5424973 \pi ^2 \zeta _3}{864}
\nonumber\\&&\mbox{}
-\frac{17813807 \zeta _3}{96}+\frac{168103 \pi ^2 \zeta _5}{180}+\frac{2973053
  \zeta _5}{480}+\frac{49508561 \zeta _7}{1344}+\frac{95485771 \pi
  ^8}{54432000}
\nonumber\\&&\mbox{}
+\frac{5863657 \pi ^6}{362880}-\frac{3526939 \pi ^4}{2880}-\frac{13246403 \pi ^2}{576}+\frac{3863977}{12}
\Bigg]\,,
%
%
\nonumber\\
\lefteqn{G_{111110001110}^{\text{(786)}}=}
\nonumber\\&&
-\frac{1}{6\epsilon^4}
%
%
-\frac{5}{3\epsilon^3}
%
%
+\frac{1}{\epsilon^2}  \Bigg[
-\frac{29}{3}-\frac{5 \pi ^2}{18}
\Bigg]
%
%
+\frac{1}{\epsilon}  \Bigg[
\frac{47 \zeta _3}{9}-\frac{31 \pi ^2}{9}-37
\Bigg]
%
%
+
\frac{326 \zeta _3}{9}-\frac{151 \pi ^4}{540}-\frac{247 \pi ^2}{9}
\nonumber\\&&\mbox{}
-49
%
%
+\epsilon  \Bigg[
\frac{361 \pi ^2 \zeta _3}{27}+\frac{170 \zeta _3}{9}+\frac{1187 \zeta _5}{15}-\frac{110 \pi ^4}{27}-\frac{539 \pi ^2}{3}+723
\Bigg]
%
%
+\epsilon^2  \Bigg[
\frac{923 \zeta _3^2}{27}
\nonumber\\&&\mbox{}
+\frac{4462 \pi ^2 \zeta _3}{27}-\frac{5846 \zeta _3}{3}+\frac{1642 \zeta _5}{3}+\frac{223 \pi ^6}{2835}-\frac{1031 \pi ^4}{27}-\frac{3167 \pi ^2}{3}+\frac{27997}{3}
\Bigg]
\nonumber\\&&\mbox{}
+\epsilon^3  \Bigg[
\frac{40046 \zeta _3^2}{27}+\frac{8161 \pi ^4 \zeta _3}{405}+\frac{35422 \pi
  ^2 \zeta _3}{27}-\frac{76862 \zeta _3}{3}+\frac{1616 \pi ^2 \zeta
  _5}{9}+\frac{3446 \zeta _5}{15}
\nonumber\\&&\mbox{}
+\frac{213683 \zeta _7}{84}-\frac{191 \pi ^6}{1260}-\frac{4408 \pi ^4}{15}-5809 \pi ^2+\frac{231793}{3}
\Bigg]
%
%
+\epsilon^4  \Bigg[
-560 s_{8 a}-\frac{25913}{81} \pi ^2 \zeta _3^2
\nonumber\\&&\mbox{}
+\frac{597926 \zeta _3^2}{27}+\frac{60682 \zeta _5 \zeta _3}{45}+\frac{119101
  \pi ^4 \zeta _3}{405}+\frac{76982 \pi ^2 \zeta _3}{9}-231202 \zeta
_3+\frac{98404 \pi ^2 \zeta _5}{45}
\nonumber\\&&\mbox{}
-\frac{152482 \zeta _5}{5}+\frac{548876 \zeta _7}{21}+\frac{2242787 \pi
  ^8}{1360800}-\frac{44237 \pi ^6}{3780}-\frac{91172 \pi ^4}{45}-\frac{275869
  \pi ^2}{9}
\nonumber\\&&\mbox{}
+\frac{1621429}{3}
\Bigg]\,,
%
%
\nonumber\\
\lefteqn{G_{111010010111}^{\text{(786)}}=}
\nonumber\\&&
\frac{1}{12\epsilon^4}
%
%
+\frac{13}{12\epsilon^3}
%
%
+\frac{1}{\epsilon^2}  \Bigg[
\frac{109}{12}+\frac{5 \pi ^2}{36}
\Bigg]
%
%
+\frac{1}{\epsilon}  \Bigg[
-\frac{73 \zeta _3}{9}+\frac{65 \pi ^2}{36}+\frac{251}{4}
\Bigg]
%
%
-\frac{895 \zeta _3}{9}-\frac{19 \pi ^4}{540}
\nonumber\\&&\mbox{}
+\frac{545 \pi ^2}{36}+\frac{1559}{4}
%
%
+\epsilon  \Bigg[
-\frac{185}{27} \pi ^2 \zeta _3-\frac{6931 \zeta _3}{9}-\frac{2896 \zeta _5}{15}-\frac{181 \pi ^4}{540}+\frac{1255 \pi ^2}{12}+\frac{9075}{4}
\Bigg]
\nonumber\\&&\mbox{}
+\epsilon^2  \Bigg[
\frac{6419 \zeta _3^2}{27}-\frac{2405 \pi ^2 \zeta _3}{27}-\frac{14309 \zeta
  _3}{3}-\frac{36133 \zeta _5}{15}-\frac{214 \pi ^6}{405}-\frac{817 \pi
  ^4}{540}+\frac{7795 \pi ^2}{12}
\nonumber\\&&\mbox{}
+\frac{151933}{12}
\Bigg]
%
%
+\epsilon^3  \Bigg[
\frac{73268 \zeta _3^2}{27}+\frac{892 \pi ^4 \zeta _3}{405}-\frac{20165 \pi ^2
  \zeta _3}{27}-\frac{76169 \zeta _3}{3}-\frac{1102 \pi ^2 \zeta _5}{9}
\nonumber\\&&\mbox{}-\frac{286879 \zeta _5}{15}-\frac{679841 \zeta _7}{168}-\frac{607 \pi ^6}{90}+\frac{137 \pi ^4}{180}+\frac{15125 \pi ^2}{4}+\frac{823921}{12}
\Bigg]
\nonumber\\&&\mbox{}
+\epsilon^4  \Bigg[
\frac{1237 s_{8 a}}{5}+\frac{13690}{81} \pi ^2 \zeta _3^2+\frac{506270 \zeta
  _3^2}{27}+\frac{422812 \zeta _5 \zeta _3}{45}+\frac{5833 \pi ^4 \zeta
  _3}{405}-\frac{46435 \pi ^2 \zeta _3}{9}
\nonumber\\&&\mbox{}-117487 \zeta _3-\frac{14119 \pi ^2 \zeta
  _5}{9}-\frac{614281 \zeta _5}{5}-\frac{4406629 \zeta _7}{84}-\frac{11991739
  \pi ^8}{6804000}-\frac{14911 \pi ^6}{270}
\nonumber\\&&\mbox{}+\frac{16381 \pi ^4}{180}+\frac{759665 \pi ^2}{36}+\frac{4378933}{12}
\Bigg]\,,
%
%
\nonumber\\
\lefteqn{G_{111010010112}^{\text{(786)}}=}
\nonumber\\&&
-\frac{1}{18\epsilon^5}
%
%
-\frac{2}{9\epsilon^4}
%
%
+\frac{1}{\epsilon^3}  \Bigg[
-\frac{2}{3}-\frac{\pi ^2}{36}
\Bigg]
%
%
+\frac{1}{\epsilon^2}  \Bigg[
\frac{116 \zeta _3}{27}-\frac{\pi ^2}{9}-\frac{16}{9}
\Bigg]
%
%
+\frac{1}{\epsilon}  \Bigg[
\frac{464 \zeta _3}{27}+\frac{8 \pi ^4}{135}-\frac{\pi ^2}{3}
\nonumber\\&&\mbox{}-\frac{40}{9}
\Bigg]
%
%
+
\frac{37 \pi ^2 \zeta _3}{27}+\frac{464 \zeta _3}{9}+\frac{3557 \zeta _5}{45}+\frac{32 \pi ^4}{135}-\frac{257 \pi ^2}{54}-\frac{32}{3}
%
\nonumber\\&&\mbox{}
+\epsilon  \Bigg[
-\frac{11323 \zeta _3^2}{81}+\frac{148 \pi ^2 \zeta _3}{27}+\frac{2458 \zeta _3}{27}+\frac{14228 \zeta _5}{45}+\frac{691 \pi ^6}{3402}+\frac{32 \pi ^4}{45}-\frac{1985 \pi ^2}{54}-\frac{224}{9}
\Bigg]
\nonumber\\&&\mbox{}
+\epsilon^2  \Bigg[
-\frac{45292 \zeta _3^2}{81}-\frac{319 \pi ^4 \zeta _3}{81}+\frac{148 \pi ^2
  \zeta _3}{9}-\frac{1910 \zeta _3}{27}+\frac{1246 \pi ^2 \zeta
  _5}{45}+\frac{14228 \zeta _5}{15}+\frac{675607 \zeta _7}{504}
\nonumber\\&&\mbox{}+\frac{1382 \pi ^6}{1701}-\frac{371 \pi ^4}{135}-\frac{4705 \pi ^2}{18}-\frac{512}{9}
\Bigg]
%
%
+\epsilon^3  \Bigg[
\frac{167 s_{8 a}}{5}-\frac{2738}{81} \pi ^2 \zeta _3^2-\frac{45292 \zeta
  _3^2}{27}
\nonumber\\&&\mbox{}-\frac{614624 \zeta _5 \zeta _3}{135}-\frac{1276 \pi ^4
  \zeta _3}{81}+\frac{19018 \pi ^2 \zeta _3}{81}-\frac{20230 \zeta
  _3}{9}+\frac{4984 \pi ^2 \zeta _5}{45}+\frac{151444 \zeta _5}{45}
\nonumber\\&&\mbox{}
+\frac{675607 \zeta _7}{126}+\frac{2968111 \pi ^8}{6804000}+\frac{1382 \pi ^6}{567}-\frac{991 \pi ^4}{27}-\frac{98941 \pi ^2}{54}-128
\Bigg]\,,
%
%
\nonumber\\
\lefteqn{G_{101111011110}^{\text{(786)}}=}
\nonumber\\&&
+\frac{1}{\epsilon^3}  \Bigg[
-\frac{1}{12} \pi ^2 \zeta _3-5 \zeta _5
\Bigg]
%
%
+\frac{1}{\epsilon^2}  \Bigg[
-\frac{37 \zeta _3^2}{2}-\frac{1793 \pi ^6}{45360}
\Bigg]
%
%
+\frac{1}{\epsilon}  \Bigg[
-\frac{197}{360} \pi ^4 \zeta _3+\frac{29 \pi ^2 \zeta _5}{6}-\frac{963 \zeta _7}{2}
\Bigg]
\nonumber\\&&\mbox{}
+ 
380 s_{8 a}+\frac{1045}{36} \pi ^2 \zeta _3^2-\frac{269 \zeta _5 \zeta _3}{6}-\frac{1047407 \pi ^8}{2721600}
\,,
%
%
\nonumber\\
\lefteqn{G_{111001111110}^{\text{(786)}}=}
\nonumber\\&&
+\frac{1}{\epsilon}  \Bigg[
2 \pi ^2 \zeta _3+10 \zeta _5
\Bigg]
%
%
+ 
10 \zeta _3^2+20 \pi ^2 \zeta _3+100 \zeta _5+\frac{176 \pi ^6}{2835}
%
\nonumber\\&&\mbox{}
+\epsilon  \Bigg[
100 \zeta _3^2+\frac{25 \pi ^4 \zeta _3}{36}+168 \pi ^2 \zeta _3-\frac{3 \pi ^2 \zeta _5}{2}+840 \zeta _5+\frac{1509 \zeta _7}{8}+\frac{352 \pi ^6}{567}
\Bigg]
\nonumber\\&&\mbox{}
+\epsilon^2  \Bigg[
-\frac{962 s_{8 a}}{5}-\frac{547}{3} \pi ^2 \zeta _3^2+840 \zeta
_3^2-\frac{4870 \zeta _5 \zeta _3}{3}+\frac{125 \pi ^4 \zeta _3}{18}+1360 \pi
^2 \zeta _3-15 \pi ^2 \zeta _5
\nonumber\\&&\mbox{}
+6800 \zeta _5+\frac{7545 \zeta _7}{4}-\frac{24163 \pi ^8}{567000}+\frac{704 \pi ^6}{135}
\Bigg]\,,
%
%
\nonumber\\
\lefteqn{G_{111101101110}^{\text{(786)}}=}
\nonumber\\&&
+\frac{1}{\epsilon^4}  \Bigg[
\frac{\pi ^2}{12}
\Bigg]
%
%
+\frac{1}{\epsilon^3}  \Bigg[
\frac{7 \zeta _3}{2}+\frac{\pi ^2}{3}
\Bigg]
%
%
+\frac{1}{\epsilon^2}  \Bigg[
14 \zeta _3+\frac{17 \pi ^4}{72}+\pi ^2
\Bigg]
\nonumber\\&&\mbox{}
+\frac{1}{\epsilon}  \Bigg[
-\frac{71}{18} \pi ^2 \zeta _3+42 \zeta _3+\frac{133 \zeta _5}{2}+\frac{17 \pi ^4}{18}+\frac{8 \pi ^2}{3}
\Bigg]
\nonumber\\&&\mbox{}
%
-\frac{1429 \zeta _3^2}{6}-\frac{142 \pi ^2 \zeta _3}{9}+112 \zeta _3+266 \zeta _5+\frac{1927 \pi ^6}{7560}+\frac{17 \pi ^4}{6}+\frac{20 \pi ^2}{3}
%
\nonumber\\&&\mbox{}
+\epsilon  \Bigg[
-\frac{2858 \zeta _3^2}{3}-\frac{8879 \pi ^4 \zeta _3}{540}-\frac{142 \pi ^2
  \zeta _3}{3}+280 \zeta _3-\frac{489 \pi ^2 \zeta _5}{10}+798 \zeta
_5+\frac{395 \zeta _7}{4}+\frac{1927 \pi ^6}{1890}
\nonumber\\&&\mbox{}
+\frac{68 \pi ^4}{9}+16 \pi ^2
\Bigg]
%
%
+\epsilon^2  \Bigg[
\frac{2582 s_{8 a}}{5}+\frac{6085}{54} \pi ^2 \zeta _3^2-2858 \zeta
_3^2-\frac{126577 \zeta _5 \zeta _3}{15}-\frac{8879 \pi ^4 \zeta _3}{135}
\nonumber\\&&\mbox{}
-\frac{1136 \pi ^2 \zeta _3}{9}+672 \zeta _3-\frac{978 \pi ^2 \zeta
  _5}{5}+2128 \zeta _5+395 \zeta _7-\frac{1170293 \pi ^8}{2268000}+\frac{1927
  \pi ^6}{630}+\frac{170 \pi ^4}{9}
\nonumber\\&&\mbox{}+\frac{112 \pi ^2}{3}
\Bigg]\,,
%
%
\nonumber\\
\lefteqn{G_{111111001110}^{\text{(786)}}=}
\nonumber\\&&
+\frac{1}{\epsilon^3}  \Bigg[
6 \zeta _3
\Bigg]
%
%
+\frac{1}{\epsilon^2}  \Bigg[
60 \zeta _3+\frac{2 \pi ^4}{15}
\Bigg]
%
%
+\frac{1}{\epsilon}  \Bigg[
-6 \pi ^2 \zeta _3+504 \zeta _3+78 \zeta _5+\frac{4 \pi ^4}{3}
\Bigg]
\nonumber\\&&\mbox{}
+ \Bigg[
-632 \zeta _3^2-60 \pi ^2 \zeta _3+4080 \zeta _3+780 \zeta _5-\frac{38 \pi ^6}{135}+\frac{56 \pi ^4}{5}
\Bigg]
%
%
+\epsilon  \Bigg[
-6320 \zeta _3^2-\frac{1343 \pi ^4 \zeta _3}{45}
\nonumber\\&&\mbox{}-504 \pi ^2 \zeta _3+32736 \zeta _3+2 \pi ^2 \zeta _5+6552 \zeta _5-4476 \zeta _7-\frac{76 \pi ^6}{27}+\frac{272 \pi ^4}{3}
\Bigg]
\nonumber\\&&\mbox{}
+\epsilon^2  \Bigg[
5088 s_{8 a}+600 \pi ^2 \zeta _3^2-53088 \zeta _3^2-\frac{84184 \zeta _5 \zeta
  _3}{5}-\frac{2686 \pi ^4 \zeta _3}{9}-4080 \pi ^2 \zeta _3+262080 \zeta _3
\nonumber\\&&\mbox{}
+20 \pi ^2 \zeta _5+53040 \zeta _5-44760 \zeta _7-\frac{76469 \pi ^8}{14175}-\frac{1064 \pi ^6}{45}+\frac{10912 \pi ^4}{15}
\Bigg]\,,
%
%
\nonumber\\
\lefteqn{G_{111011011110}^{\text{(786)}}=}
\nonumber\\&&
+\frac{1}{\epsilon}  \Bigg[
\frac{7 \pi ^4 \zeta _3}{360}-\frac{5 \pi ^2 \zeta _5}{3}-\frac{441 \zeta _7}{16}
\Bigg]
%
%
+ \Bigg[
-\frac{87 s_{8 a}}{2}-\frac{23}{6} \pi ^2 \zeta _3^2-\frac{473 \zeta _5 \zeta _3}{2}-\frac{8069 \pi ^8}{777600}
\Bigg]\,,
%
%
\nonumber\\
\lefteqn{G_{111011011120}^{\text{(786)}}=}
\nonumber\\&&
+\frac{1}{\epsilon^5}  \Bigg[
-\frac{\pi ^2}{96}
\Bigg]
%
%
+\frac{1}{\epsilon^4}  \Bigg[
\frac{7 \zeta _3}{16}
\Bigg]
%
%
+\frac{1}{\epsilon^3}  \Bigg[
\frac{463 \pi ^4}{8640}
\Bigg]
%
%
+\frac{1}{\epsilon^2}  \Bigg[
\frac{1247 \zeta _5}{48}-\frac{127 \pi ^2 \zeta _3}{144}
\Bigg]
\nonumber\\&&\mbox{}
+\frac{1}{\epsilon}  \Bigg[
\frac{38761 \pi ^6}{362880}-\frac{1079 \zeta _3^2}{12}
\Bigg]
%
%
-\frac{78923 \pi ^4 \zeta _3}{12960}+\frac{18301 \pi ^2 \zeta _5}{720}-\frac{161 \zeta _7}{24}
%
\nonumber\\&&\mbox{}
+\epsilon  \Bigg[
\frac{3291 s_{8 a}}{5}+\frac{55183}{216} \pi ^2 \zeta _3^2-\frac{330689 \zeta _5 \zeta _3}{90}-\frac{122374187 \pi ^8}{653184000}
\Bigg]\,.
%
%
\end{eqnarray}

\end{appendix}


\end{document}